\documentclass[12pt]{article}
\usepackage{multirow, amsmath, titling, breqn}
 \usepackage[colorlinks]{hyperref}
 \hypersetup{
 colorlinks=true,
 linkcolor=blue,
 filecolor=magenta, 
 urlcolor=blue,
}
\usepackage{comment}
\usepackage[inline]{enumitem}
\usepackage{tgtermes, hyperref} 
\usepackage[table]{xcolor} 
\usepackage{multicol,breqn, bm,tikz,tabularx,ragged2e,booktabs,caption,float, lscape, lipsum, setspace}
\usepackage{lscape,afterpage}
\usepackage[super,comma]{natbib}
\usepackage[margin = 1in]{geometry}

\begin{document}

\title{Surrogacy Validation for Time-to-Event Outcomes with Illness-Death Frailty Models}
\author{Emily K. Roberts$^{1*}$, Michael R. Elliott$^{2,3}$, Jeremy M. G. Taylor$^2$\\
$^1$Department of Biostatistics, University of Iowa, Iowa City, IA\\
$^2$Department of Biostatistics, University of Michigan, Ann Arbor, MI\\
$^3$Survey Methodology Program, Institute for Social Research Ann Arbor, MI\\
$^{*}$145 N. Riverside Dr. Iowa City, IA 52242\\
$^{*}$emily-roberts-1@uiowa.edu
}
\date{}
\maketitle

\section*{Key Words:} Bayesian methods, clinical trial, illness death model, surrogacy validation, time-to-event data

\section*{Abstract}

A common practice in clinical trials is to evaluate a treatment effect on an intermediate endpoint when the true outcome of interest would be difficult or costly to measure. We consider how to validate intermediate endpoints in a causally-valid way when the trial outcomes are time-to-event. Using counterfactual outcomes, those that would be observed if the counterfactual treatment had been given, the causal association paradigm assesses the relationship of the treatment effect on the surrogate $S$ with the treatment effect on the true endpoint $T$. In particular, we propose illness death models to accommodate the censored and semi-competing risk structure of survival data. The proposed causal version of these models involves estimable and counterfactual frailty terms. Via these multi-state models, we characterize what a valid surrogate would look like using a causal effect predictiveness plot. We evaluate the estimation properties of a Bayesian method using Markov Chain Monte Carlo and assess the sensitivity of our model assumptions. Our motivating data source is a localized prostate cancer clinical trial where the two survival endpoints are time to distant metastasis and time to death.

\newpage \section{Introduction} 

Time-to-event endpoints are common in oncology trials, though it can often take many years to accrue enough observed events to complete the study (Kemp et al. 2017). In a randomized clinical trial, an appropriate surrogate endpoint can serve as a substitute indicator for if a treatment effect exists on some true outcome of interest. In this work, our data come from a prostate cancer clinical trial with a binary treatment {of adding anti-androgen therapy to an existing regimen} (Shipley et al. 2017). The two endpoints of interest are the occurrence of distant metastasis and overall survival. Here the terminal event is death from any cause and is the primary endpoint for the trial. For these patients, death from prostate cancer will only occur if the person has had metastases. However, some men will experience death during follow-up with or without experiencing distant metastases spreading first. Overall survival is therefore a mixture of two death types, death from prostate cancer and death from other causes. However, in the data the cause of death may not be known. Mechanistically understanding whether distant metastases is a desirable surrogate for overall survival in this setting may be beneficial for clinicians and trialists.

Given the substantial risk of potentially using an invalid surrogate endpoint in a large-scale trial, rigorous standards have been proposed to validate a surrogate (Vanderweele, 2013). The first criteria to determine the validity of candidate surrogate endpoints were suggested by Prentice (1989) which test whether a treatment affects the true endpoint only through the pathway of the surrogate endpoint. While the criteria are applicable to {different outcomes such as} time-to-event endpoints {that we will be focusing on}, they involve regression models that rely on conditioning on the observed value of $S$, leading to a non-causal interpretation. More recent frameworks to determine if a surrogate marker is appropriate for use in a future trial can be broadly grouped into the causal effects and causal association paradigms (Joffe and Greene, 2009). The causal association framework aims to evaluate the relationship of the treatment effect on the surrogate $S$ with the treatment effect on the true clinical endpoint $T$. These methods are often built upon counterfactual outcomes $T(z)$, which are the clinical outcomes of interest, and $S(z)$, the surrogate endpoints, where the notation $Z = z$ represents treatment under either the observed or counterfactual assignment.

Methods within the causal association framework have been proposed for trials where the true outcome $T$ is a time-to-event outcome under different corresponding surrogate endpoint types. Tanaka et al. (2017) consider a binary surrogate for a survival primary outcome within the meta-analytic framework, and Gao (2012) considers a time-to-event $T$ and binary $S$ for a single trial using principal stratification methods (Frangakis and Rubin, 2002). Taylor et al. (2015) propose a Gaussian copula model with a survival endpoint for $T$ and ordinal endpoint $S$. The principal stratification estimand proposed by Qin et al. (2008) allows for a continuous $S$ and time-to-event $T$. This was expanded upon in Gabriel and Gilbert (2014) and Gabriel, Sachs, and Gilbert (2015) in pursuit of a causal effect interpretation. Causal solutions for validation become more challenging when the surrogate is also subject to censoring. Instead, others such as Parast and colleagues (2017) rely on different measures such as proportion explained for time-to-event outcomes, and likewise Hsu et al. (2015), Vandenberghe et al. (2018), and Weir et al. (2021) address time-varying surrogates using mediation approaches that rely on proportion mediated metrics within the causal effects paradigm.

To our knowledge, the setting where both $S$ and $T$ are time-to-event endpoints has not been fully addressed within the principal stratification framework. Building on the work of Frangakis and Rubin (2002), we aim to develop a corresponding Causal Effect Predictiveness (CEP) curve proposed by Gilbert and Hudgens (2008) to validate a surrogate endpoint when both $S$ and $T$ are time-to-event. The key to obtaining a causal assessment in this paradigm is classifying individuals based on their set of potential values of the post-treatment variable, which here would be the surrogate endpoint. In a simple case where $S$ and $T$ are Gaussian outcomes and $Z$ takes on the value 0 or 1, the analog to surrogate-specific strata and the corresponding CEP curve for validation is based on the quantity $E(T(1)-T(0)|S(1)-S(0) = s).$ Briefly, the CEP criteria intuitively assert that there be no average treatment effect on $T$ for the strata of patients defined by no treatment effect on $S$, and conversely that there exist an overall treatment effect on $T$ for the strata of patients defined by a treatment effect on $S$. A comparable contrast and consideration of principal strata when $T(z)$ and $S(z)$ are subject to censoring and a semi-competing risk structure will be explored in this paper.

Outside of the surrogacy validation setting, semi-competing risks based on counterfactual hazards have been explored (Huang, 2021). Within the principal stratification framework, unobserved outcomes due to truncation by death can be addressed by defining strata based on survivorship cohorts (Zhang and Rubin, 2003). Comment et al. (2019) define a survivor average causal effect in the presence of a semi-competing risk where principal causal effects are defined for individuals who would survive regardless of the assigned treatment. Xu et al. (2020) propose a causal estimand for a semi-competing risk structure to address truncation by death $\frac{P(S(1) < \tau | T(0) \geq \tau, T(1) \geq \tau)}{P(S(0) < \tau | T(0) \geq \tau, T(1) \geq \tau)}$ which conditions on these survivor principal strata.

The estimands for surrogacy validation with a continuous $S$ by Qin et al. (2008) and Gabriel, Sachs, and Gilbert (2015) described earlier can be written as $1 - \frac{P(T(1) = \tau | T(1) \geq \tau_{k-1}, S(1) = s_1, S(0) = s_0)}{P(T(0) = \tau | T(0) \geq \tau_{k-1}, S(1) = s_1, S(0) = s_0)}$ and $\frac{1-P(T(1) > t | T(0) \geq \tau, T(1) \geq \tau, S(1) = s_1, S(0) = s_0)}{1-P(T(0) > t | T(0) \geq \tau, T(1) \geq \tau, S(1) = s_1, S(0) = s_0)}$ for some time $\tau$, respectively. Whereas the previous CEP quantities suggest conditioning on counterfactual surrogate outcomes, this becomes less straightforward in our setting. While existing models are suitable for these data that account for semi-competing risks, $S$ may not be well-defined if it is not observed before $T$. The proper corresponding surrogacy validation estimand is less readily apparent since it may not be possible to condition on strata defined by $S(0)$ and $S(1)$ occurring or not by time $\tau$ (see a discussion regarding estimands in Buhler et al. 2022). For example, while we can construct $\frac{P(T(1) < \tau | S(0) \geq \tau, S(1) \geq \tau)}{P(T(0) < \tau | S(0) \geq \tau, S(1) \geq \tau)}$ or $\frac{P(T(1) < \tau | T(0) \geq \tau, S(0) \geq \tau, S(1) \geq \tau)}{P(T(0) < \tau | T(1) \geq \tau, S(0) \geq \tau, S(1) \geq \tau)}$ for some time $\tau$, it is not clear which would be a principled estimand to use for validation with our endpoint types. 

Rather than conditioning on surrogate outcomes, we develop a principal stratification approach that conditions on counterfactual hazards and outline causal quantities based on these. We propose an illness-death model to incorporate the censored and semi-competing risk structure of the data. Previous work using principal surrogacy for repeated outcome measurements incorporates estimation of subject-specific random effects (Roberts et al., 2022). Here we utilize frailty terms to capture subject specific heterogeneity and allow dependence among the transitions of the illness-death model. Frailties have been proposed for surrogate validation settings that differ from our single trial with subject-level, counterfactual outcomes. These methods include joint frailty-copula models for meta-analysis to define valid surrogates (Emura et al., 2017; Sofeau, Emura, and Rondeau, 2019; Sofeau, Emura, and Rondeau, 2020). 

In Section 2, we propose the causal modeling strategy based on the illness-death approach for a single trial and {link} this formulation to the Prentice criteria. In Section 3, we provide the likelihood of the illness-death model and propose a Bayesian estimation strategy. Section 4 describes our proposed CEP quantities and explores {CEP plots that correspond to different data settings to help define} what an ideal surrogate would look like. A simulation study is provided in Section 5 with a real data analysis from a prostate cancer trial in Section 6. Discussion and future work are provided in Section 7.

\section{Illness-Death Approach}

The structure of the illness-death model is a natural way to describe data with the semi-competing risk structure and has potential use for surrogacy validation (O'Quigley and Flandre, 2012). Here we consider counterfactual illness-death models and the principal stratification framework. Let $T_{jk}(z)$ denote the gap time between two states $(j = 1,2, k = 2,3)$ and corresponding transition intensities $\lambda_{jk}^z$ between states in the treatment-specific illness-death models for treatment $Z= z$ as shown in Figure \ref{fig:illnessdeath}.

Notably, this conceptualization is related to the models used in the Prentice criteria (1989). In short, the Prentice criteria assess whether a) the treatment and true endpoint are conditionally independent, given the surrogate endpoint, and b) the surrogate and the treatment are correlated. This determination is made by fitting two regression models and determining if the coefficient for the treatment effect on $T$ becomes null after adjusting for the surrogate in the model. These ensure that a treatment effect on the true endpoint will imply a treatment effect on the surrogate endpoint. In particular, Prentice's measures, which identify statistical surrogates, are only correlative. 

We propose a more rigorous and flexible strategy to identify a consistent surrogate using potential outcomes and counterfactual illness-death models in pursuit of a causal interpretation (VanderWeele, 2013). Motivation for our proposed models can be seen through a special case of regression models that are related to models used to evaluate the Prentice criteria. In the Prentice criteria, we can consider three models:

 For time to $S$, A) $\lambda(t) \exp(\phi_0 Z_i + \eta_0 X_i)$
 
 For time to $T$, B) $\lambda(t) \exp(\phi_1 Z_i + \eta_1 X_i)$ and C) $\lambda(t) \exp(\phi_2 Z_i + \eta_2 X_i + \omega I(t>S_i))$
 
 \noindent where $S$ denotes the time of the surrogate outcome occurring, $Z$ denotes treatment, $X$ denotes baseline covariates, and time $t$ is measured from randomization. Then the difference in the $\phi_1$ and $\phi_2$ coefficients between B and C largely captures the value of the surrogate. In comparison, consider a general set of observed data models for the three transitions \begin{equation}
\lambda_{12}(t)\exp(\omega_{12i} + \phi_3 \ Z_i + \eta_3 \ X_i)
\end{equation}\vspace{-1.3cm}
$$\lambda_{13}(t)\exp(\omega_{13i} + \phi_4 \ Z_i + \eta_4 \ X_i)$$
$$\lambda_{23}(t)\exp( \omega_{23i} + \theta \ S_i + \phi_5 \ Z_i + \eta_5 \ X_i + \beta \ S_i \ Z_i )$$

\noindent where $\omega_{jk}$ denote frailty terms. Our proposed models include a model following the occurrence of $S$, allow for more interaction terms, and include frailties. Further extension of the models and their connection with the counterfactual illness-death models in Figure \ref{fig:illnessdeath} can be found in an appendix. In the model we propose and explore in detail in the following sections, each counterfactual arm has its own set of transition hazard models. We will first consider all counterfactual quantities that appear in the complete data likelihood for the proposed model.

\subsection{Defining Causal Quantities Based on Hazards and Frailty Models}

We propose to model the transition hazards that correspond to the gap times $T_{jk}(z)$ in Figure \ref{fig:illnessdeath}. Shared or common frailty terms, which quantify the dependence between the different processes within the same person, can provide information on the dependence structure between the time to intermediate event and the time to terminating event in multi-state models (Zhang et al., 2014; Xu et al., 2010). In models for time-to-event data frailties are commonly incorporated to model correlation among events, to allow for heterogeneity among individuals, or to capture the effect of some omitted covariate. In our setting, we consider both counterfactual outcomes and transitions, and we want to allow for possible dependence between the counterfactual outcomes. As this association is integral to the value of the surrogate, we propose to use illness-death frailty models where the hazards are linked via frailty terms. Here we consider multiple hazards with frailties both to allow dependence across state transitions and to link observable transitions in arm $Z=z$ to the counterfactual transitions for $Z = 1-z$. 

For a single time-to-event and a general frailty $\omega$, the hazard can be written $\lambda(t|X, \beta, \omega, \kappa) = \lambda_{0}(t)\exp( \kappa \omega + X\beta)$, where $\omega$ has some pre-specified distribution and may have an associated coefficient parameter $\kappa$. Various assumptions can be made about the frailty term $\omega$, such as that it follows a Normal or Gamma distribution, for simplicity and computational feasibility. For the illness-death models specified in Figure \ref{fig:illnessdeath}, a set of the six correlated frailties are required, one for each model. However, for identifiability and computational concerns, we impose some restrictions and simplifying assumptions. We initially propose two different formulations of the sets of models, and for ease of notation, we exclude baseline covariates $X$. 

\subsection*{Model A {using Time Dependent Covariates}}

For $z=0$,
\begin{equation}
 \lambda_{12}^0(t | \omega_{12i}^0) = \lambda_{12,0}^0(t) \exp(\kappa_{12}^0 \omega_{12i}^0)\end{equation}
 $$\lambda_{13}^0(t | \omega_{13i}^0) = \lambda_{13,0}^0(t) \exp(\kappa_{13}^0 \omega_{13i}^0 )$$
$$\lambda_{23}^0(t |T_{12i}(0), \omega_{23i}^0) = \lambda_{23,0}^0(t-T_{12i}(0)) \exp(\kappa_{23}^0 \omega_{23i}^0 + \theta_{23}^0 T_{12i}(0))I(t > T_{12i}(0))$$
Similarly for $z = 1,$
$$\lambda_{12}^1(t | \omega_{12i}^1) = \lambda_{12,0}^1(t) \exp(\kappa_{12}^1 \omega_{12i}^1)$$
$$\lambda_{13}^1(t | \omega_{13i}^1) = \lambda_{13,0}^1(t) \exp(\kappa_{13}^1 \omega_{13i}^1 )$$
$$\lambda_{23}^1(t | T_{12i}(1), \omega_{23i}^1) = \lambda_{23,0}^1(t-T_{12i}(1)) \exp(\kappa_{23}^1 \omega_{23i}^1 + \theta_{23}^1 T_{12i}(1))I(t > T_{12i}(1))$$ 

\noindent where $T_{12i}$ is the time that subject $i$ moves into state $S$. We include $\theta_{23}$ in the $\lambda_{23}$ model as the coefficient for our time dependent covariate $T_{12}$. The purpose is to capture the effect of this transition time, and the time that an individual experiences $S$ may help to assess the strength of association between $S$ and $T$. We model the transition using a clock reset for $\lambda_{23}$ (ie the time scale is $t-T_{12}(z)$). 

The restrictions and assumptions we will be considering are to make $\omega_{13i}^z = \omega_{23i}^z$ and to set some of the $\kappa_{jk}^z = 1.$ If the $\kappa$ parameters vary, they essentially influence how variable the frailty terms are. We will refer to $\kappa$ as frailty coefficients. One rationale for assuming $\omega_{13i}^z = \omega_{23i}^z$ in this setting is that both are frailties that influence time to death from other causes in our motivating trial. For example, since our variable $T$ is death from any cause, we may expect that some men will die of old age. It may be reasonable to expect that an individual may have their own propensity for experiencing death from other causes irrespective of whether or not $S$ has occurred. Another consideration is that by including the coefficient for our time-varying covariate, $\theta_{23}^z$, the model captures the magnitude of the effect for the time it takes to experience the intermediate outcome $S$. This makes it more plausible that certain frailties are equal and conditional independence assumptions may be more likely. Lastly, the frailties capture heterogeneity on the individual level. There may still be heterogeneity on the population level for the variability in the hazard of going from baseline to $T$ or from $S$ to $T$ which can be reflected in the baseline hazards. We explore these variations in later sections.

\subsection*{Model B using Multiple Frailties in Place of Time Dependent Covariates}

We include an alternate option to incorporate the dependence between the different transitions such as a model that includes two frailty terms in the $S \rightarrow T$ transition
\begin{equation}
 \lambda_{12}^0(t | \omega_{12i}^0) = \lambda_{12,0}^0(t) \exp(\kappa_{12}^0 \omega_{12i}^0)\end{equation}
 \vspace{-1.3cm}
 $$\lambda_{13}^0(t | \omega_{13i}^0) = \lambda_{13,0}^0(t) \exp(\kappa_{13}^0 \omega_{13i}^0 )$$
$$\lambda_{23}^0(t | T_{12i}(0),\omega_{13i}^{*0}, \omega_{12i}^{*0}) = \lambda_{23,0}^0(t-T_{12i}(0)) \exp(\kappa_{12}^{*0} \omega_{12i}^0 + \kappa_{13}^{*0} \omega_{13i}^0)I(t > T_{12i}(0))$$
$$\lambda_{12}^1(t | \omega_{12i}^1) = \lambda_{12,0}^1(t) \exp(\kappa_{12}^1 \omega_{12i}^1)$$ $$\lambda_{13}^1(t | \omega_{13i}^1) = \lambda_{13,0}^1(t) \exp(\kappa_{13}^1 \omega_{13i}^1 )$$
$$\lambda_{23}^1(t | T_{12i}(1),\omega_{13i}^{*1}, \omega_{12i}^{*1}) = \lambda_{23,0}^1(t-T_{12i}(1)) \exp(\kappa_{12}^{*1} \omega_{12i}^1 + \kappa_{13}^{*1} \omega_{13i}^1)I(t > T_{12i}(1))$$

\noindent The motivation of this model is an alternative way to capture the subject specific relationship between the different transitions via the $\kappa_{12}^{*}$ and $\kappa_{13}^{*}$ coefficients. This model does not include $T_{12}$ as a time-varying covariate. When we assume $\omega_{23}^z = \omega_{13}^z,$ the key difference between models A and B is the way in which the transition from baseline to the intermediate outcome and the time following that transition are related; these are linked using either a time varying covariate (in model A) or another frailty term (in model B). Again, the frailty coefficients $\kappa$ can be thought of parameters that increase or decrease the magnitude of the effect of the frailties. We would not expect $\kappa_{12}^{*z}$ and $\kappa_{12}^z$ to be necessarily equal across the models given the different assumptions in each model.  

\subsection*{Frailty Structures}

In its most generality, model A has six correlated frailties, which we assume have a multivariate normal distribution.

{$$\left( \begin{array}{c} \omega_{12i}^0\\ \omega_{12i}^1 \\ \omega_{13i}^0 \\ \omega_{13i}^1 \\ \omega_{23i}^0 \\ \omega_{23i}^1 \end{array}
\right) \sim N \left( \left( \begin{array}{c} 0 \\ 0 \\ 0 \\ 0 \\ 0 \\ 0 \end{array} \right), 
\left( \begin{array}{cccccc} 1 & \rho_S & \rho_{00} & \rho_{01} & \rho_{S1} & \rho_{S2} \\ & 1 & \rho_{10} & \rho_{11} & \rho_{S3} & \rho_{S4} \\
&& 1 & \rho_T & \rho_{T1} & \rho_{T2} \\
&& & 1 & \rho_{T3} & \rho_{T4} \\
&&&& 1 & \rho_{ST} 
\\ 
&&&&& 1 \\ \end{array} \right) \right)
$$}

While this model has a very general form, it may not be necessary or even desirable to consider this level of generality. We will be focusing on special cases of this general model, which we think are appropriate for the setting of surrogacy assessment.

To reduce the number of frailties to estimate to four in model A, we assume that both transitions into $T$ have the same frailty ($\omega_{13}^z = \omega_{23}^z$) since they are both relevant for time to the terminal event. As discussed above, since the terminal event is death from any cause, it seems justifiable to assume that conditional on all other terms in the model, frailties toward death from any cause would be the same on the individual level with or without the occurrence of the intermediate event $S$. This assumption will be useful for estimation since $T_{23i}$ is not defined for all individuals. With this assumption, our transition models from $S$ to $T$ in model A can be written 

\vspace{-.56in}

$$\lambda_{23}^0(t | T_{12i}(0), \omega_{13i}^0) = \lambda_{23,0}^0(t-T_{12i}(0)) \exp(\kappa_{23}^0 \omega_{13i}^0 + \theta_{23}^0 T_{12i}(0))I(t > T_{12i}(0))$$ $$\lambda_{23}^1(t | T_{12i}(1), \omega_{13i}^1) = \lambda_{23,0}^1(t-T_{12i}(1)) \exp(\kappa_{23}^1 \omega_{13i}^1 + \theta_{23}^1 T_{12i}(1))I(t > T_{12i}(1))$$ Ideally, we could allow $\kappa_{23}^z$ to take on different values from $\kappa_{13}^z$ to accommodate different amounts of dependence between the transitions. For both models A and B we consider the joint distribution 
$$\left( \begin{array}{c} \omega_{12i}^0\\ \omega_{12i}^1 \\ \omega_{13i}^0 \\ \omega_{13i}^1 \end{array}
\right) \sim N \left( \left( \begin{array}{c} 0 \\ 0 \\ 0 \\ 0 \end{array} \right), 
\left( \begin{array}{cccc} 1 & \rho_S & \rho_{00} & \rho_{01} \\ & 1 & \rho_{10} & \rho_{11} \\
&& 1 & \rho_T \\
&& & 1 \\ 
\end{array} \right) \right)
$$

{In most of the work presented here, we will also assume} $\omega_{12i}^z \perp \omega_{13i}^z$ (the frailties for an individual are independent across states), meaning $\rho_{00} = \rho_{01} = \rho_{11} = \rho_{10} = 0$. We thus assume
\begin{equation}
\left( \begin{array}{c} \omega_{12i}^0 \\ \omega_{12i}^1 \end{array}
\right) \sim N \left( \left( \begin{array}{c} 0 \\ 0 \\ \end{array} \right), 
\left( \begin{array}{cc} 1 & \rho_S \\ \rho_S & 1 \\ \end{array} \right) \right) \ \ \ \ \ \ \ \ \left( \begin{array}{c} \omega_{13i}^0 \\ \omega_{13i}^1 \end{array}
\right) \sim N \left( \left( \begin{array}{c} 0 \\ 0 \\ \end{array} \right), 
\left( \begin{array}{cc} 1 & \rho_T \\ \rho_T & 1 \\ \end{array} \right) \right)
\end{equation}

\noindent This type of assumption may aid in estimation. We could instead impose a strong assumption of shared frailties for each arm: $\omega_{12i}^0 =\omega_{13i}^0=\omega_{23i}^0$ and $\omega_{12i}^1 =\omega_{13i}^1=\omega_{23i}^1$. The motivation for this comes from considering the frailty as representing an omitted covariate. We do not further pursue this assumption. 

\subsection{Identifiability and Sensitivity Analysis}

Certain parameters within our model are non-identifiable because they describe relationships between counterfactual variables ($\rho_S$ and $\rho_T$ for example), while others are ``barely" identifiable (the combination of the baseline hazard, frailties, and the $\kappa$ parameters, for example) and are therefore hard to estimate. In particular, the frailty terms are weakly identified based on which events, $S$ and/or $T$, are actually observed. Since we have made modeling assumptions to aid in estimation, we can evaluate the sensitivity of the assumed models in several ways. Because the parameters $\rho_S$ and $\rho_T$ in the complete data likelihood are not identifiable, they will be fixed at preset values in our proposed method (and later we will discuss if the complete data likelihood is necessary). Based on biological considerations under the counterfactual framework, we may not expect these correlation parameters to be negative or exactly equal to one. Further, we can vary which frailties are assumed to be independent or equal, alter which values of $\kappa_{jk}^z$ are set to one, change the baseline hazard from a Weibull distribution to piecewise exponential or something more flexible, assess different effects of covariates in the transitions, and modify our proposed time-reset parameterization. We provide a tool for assessing the sensitivity of these values and commentary on the feasibility and identifiability of estimating these models with and without these assumptions in later sections.

\section{Likelihood and Estimation}

\subsection{Likelihood Contributions}

We consider a randomized clinical trial of $n$ subjects for a binary treatment $Z$. For generality, let $n_z$ denote the number of subjects in treatment arm $Z = z$ (and we may assume that $n/2$ subjects are in treatment group $z = 1$ and $n/2$ are in treatment group $z=0$ since the treatment assignment is randomized and under the control of the investigator). Let $\{S_i, \delta_{Si}, T_i, \delta_{Ti}, X_i, Z_i \}$ be the observed data for subject $i$ for $i = 1,...,n$. We will also consider a random or administrative censoring time $C_i$. $S_i$ denotes the time to transition to state $S$, $T_i$ denotes the time that the terminal event $T$ occurs, and $\delta_T$ and $\delta_S$ denote the censoring indicators for $T$ and $S$ being observed. Then $\delta_{Ti} = 1$ when $T_i < C_i$ and $\delta_{Si} = 1$ when $S_i < C_i$ and $S_i < T_i$. 

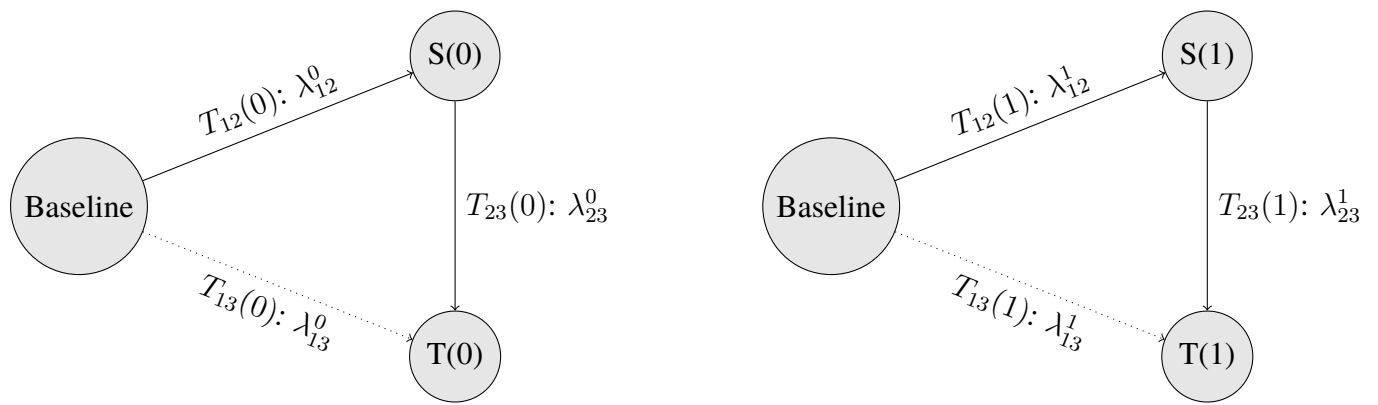
\begin{figure}[H]
\begin{tikzpicture}

 \tikzstyle{leaf}=[shape=circle,draw=black,fill=gray!20,minimum size=0.01cm]
	\node[leaf] (p3) at ( 2, -2) {T(0)};
 	\node[leaf] (p1) at (-3, 0) {Baseline};
 \node[leaf] (p2) at ( 2, 2) {S(0)};
 \begin{scope}[every path/.style={->}]
 		\draw (p1) to node[midway, above, sloped] {$T_{12}(0)$: $\lambda_{12}^0$} (p2);
 \draw[dotted] (p1) to node[midway,below, sloped] {$T_{13}(0)$: $\lambda_{13}^0$} (p3) ;
 \draw (p2) to node[right,midway] {$T_{23}(0)$: $\lambda_{23}^0$}(p3) ;
 \end{scope} 

 \tikzstyle{leaf}=[shape=circle,draw=black,fill=gray!20,minimum size=0.01cm]
	\node[leaf] (p3) at ( 12, -2) {T(1)};
 	\node[leaf] (p1) at (7, 0) {Baseline};
 \node[leaf] (p2) at ( 12, 2) {S(1)};
 \begin{scope}[every path/.style={->}]
 		\draw (p1) to node[midway, above, sloped] {$T_{12}(1)$: $\lambda_{12}^1$} (p2);
 \draw[dotted] (p1) to node[midway,below, sloped] {$T_{13}(1)$: $\lambda_{13}^1$} (p3) ;
 \draw (p2) to node[right,midway] {$T_{23}(1)$: $\lambda_{23}^1$}(p3) ;
 \end{scope} 
\end{tikzpicture}
 \caption{Counterfactual illness-death models for baseline, illness ($S$), and death ($T$). The potential pathways are labeled with the gap time and corresponding transition intensity for each treatment arm.}
 \label{fig:illnessdeath}
\end{figure}

We can also conceptualize the data in terms of the random variables in Figure \ref{fig:illnessdeath}. Based on gap times between states $T_{jk}^z$, the data can also be represented as $\{T_{12i}, T_{13i}, T_{23i}, \delta_{Si}, \delta_{Ti}, X_i, Z_i \}$, with $T_{23i}$ not defined when $S_i$ is not observed. In the illness-death formulation, there are four possible combinations of observable $\delta_{Si}$ and $\delta_{Ti}$. We assume that when neither event is observed, meaning $\delta_{Si} = \delta_{Ti} = 0,$ then $T_{12i}(z)$ and $T_{13i}(z)$ take on the same value as being censored at $C_i$. Consider when $T$ is observed before $S$, meaning $\delta_{Ti} = 1, \delta_{Si} = 0$. Then the observed data related to $S_i$ for individual $i$ is equal to $\{T_{13i}, \delta_{Si} = 0 \}$, and observed $T_i$ is based on $\{T_{13i}, \delta_{Ti} = 1 \}$, while $T_{23i}$ is not defined. Now consider when only $S$ is observed, meaning $\delta_{Ti} = 0, \delta_{Si} = 1$. Then the observed data for individual $i$ is $S_i$ based on $\{ T_{12i}, \delta_{Si} = 1\} $. Assuming $T$ is not observed after, the value $T_i$ takes on is censored at $\{ C_i, \delta_{Ti} = 0 \}$. If both $S$ and $T$ are observed with $\delta_{Ti} = \delta_{Si} = 1$, then $S_i$ is based on $\{ T_{12i}, \delta_{Si} = 1\},$ and $T_i$ is based on $\{ T_{12i} + T_{23i}, \delta_{Ti} = 1 \}.$ We provide the likelihood under these scenarios next.

We assume that each hazard in Figure \ref{fig:illnessdeath} follows a Weibull distribution, so $T_{12}(z) \sim$ $Weibull(\alpha_{12}^z,$ $\gamma_{12}^z), T_{13}(z) \sim Weibull(\alpha_{13}^z, \gamma_{13}^z),$ and $T_{23}(z) \sim Weibull(\alpha_{23}^z, \gamma_{23}^z)$ for shape parameters $\alpha_{jk}^z$ and scale parameters $\gamma_{jk}^z$. The scale and shape parameters must be positive: $\gamma_{jk}^z>0, \alpha_{jk}^z>0$. We parameterize the cumulative baseline hazard function as $\Lambda_{jk0}^z(t) = \gamma_{jk}^z t^{\alpha_{jk}^z} = \int_0^t \lambda_{jk0}^z(u)du$ for a given Weibull model, where $\lambda_{jk0}^z(t) = \gamma_{jk}^z\alpha_{jk}^z t^{\alpha_{jk}^z-1}$ and $\lambda_{jk}^z(t) = \lambda_{jk0}^z(t)\exp(\kappa_{jk}^z\omega_{jk}^z)$ for $jk$ = 12 or 13. The model for $\lambda_{23}^z$ is more complex and depends on whether model A or B is assumed; for example, model A corresponds to $\lambda_{230}^z(t)\exp(\kappa_{23}^z\omega_{23}^z + \theta_{23}^z T_{12}(z))$. 

For estimation there are two likelihoods that could be used, either the observed data likelihood, or the complete data likelihood.
The complete data likelihood is derived using the random variables in Figure \ref{fig:illnessdeath} with both sets of counterfactual outcomes under the two treatment arms $T_{12}(0), T_{12}(1), T_{13}(0), T_{13}(1), T_{23}(0), T_{23}(1)$. This approach considers the joint model of the outcomes and involve all elements $\rho$ of the correlation matrix in equation 4. Using this specification, an imputation scheme could be proposed to fill in all missing outcomes. Any relation between the potential outcomes across treatment arms for an individual in the complete data likelihood is not identified. Based on previous exploration of methods that use either the observed or the complete data likelihood (Roberts et al. 2021), using this complete data likelihood and employing imputation is not necessary to carry out the validation procedure. Here we will only focus on the observed data likelihood during estimation and consider each arm of the trial separately. For ease of notation, we will drop the superscript in this section as the derivations apply to both treatment arms. Any counterfactual quantities needed for calculation of the CEP curve will be described separately in Section 4. We note that $\{T_{23i}, \omega_{23i}\}$ are not defined when $\delta_{Si} = 0$ and do not contribute to the likelihood, which is the case for either the complete data or observed data likelihood.

The likelihood contributions can be written similarly to work done by Conlon et al. (2014b). Conditional on the frailties and the other parameters the likelihood contribution for subject $i$ is, \\ $L_i = L(T_{12i}, T_{13i}, T_{23i}, \delta_{Si}, \delta_{Ti} ; \omega_{12i}, \omega_{13i}, \omega_{23i}, \gamma_{12}, \alpha_{12},\gamma_{13}, \alpha_{13},\gamma_{23}, \alpha_{23}, \theta_{23}, \kappa_{12}, \kappa_{13}, \kappa_{23})$. For those who had not experienced $S$, we are in the setting where $\delta_{Si} = 0$, $T_{12i}=T_{13i}$ and $T_{23i}$ is not defined, then $L_i = \lambda_{13}(T_{13i})^{\delta_{Ti}}\exp(-\int_0^{T_{13i}} \lambda_{13}(u)du - \int_0^{T_{13i}}\lambda_{12}(u)du)$

\noindent For those who experience $S$, and are either dead or alive, $\delta_{Si} = 1,$ and $T_{23i}$ is defined. $\delta_{Ti}$ may be equal to either 0 or 1 depending on if the terminal event is observed:

\noindent $L_i = \lambda_{12}(T_{12i})\exp(-\int_0^{T_{12i}} \lambda_{12}(u)du - \int_0^{T_{12i}} \lambda_{13}(u)du)\lambda_{23}(T_{23i}|T_{12i})^{\delta_{Ti}}\exp(-\int_0^{T_{23i}} \lambda_{23}(u|T_{12i})du)$

\subsection{Bayesian Estimation}

To facilitate estimation, we take a Bayesian approach using Markov Chain Monte Carlo (MCMC). We use prior distributions similar to those suggested in Gao et al. (2012) and Sahu et al. (1997). Regression coefficients are assumed to have a diffuse normal prior (Sahu et al. 1997). We assume a Gamma($p_1, p_2)$ prior for the scale parameters $\gamma_{jk}$ of the Weibull distribution, and we also assume a Gamma($p_3, p_4)$ prior for the shape parameters $\alpha_{jk}$ with hyperparameters $p_1 = p_2 = p_3 = p_4 = 0.1.$

Any parameters that do not have a closed-form posterior distribution ($\alpha_{jk}^z,\gamma_{jk}^z, \omega_{jk}^z, \theta_{23}^z, \kappa_{23}^z$) are drawn using a Metropolis-Hastings step (Robert and Casella, 2004). At each iteration of the MCMC, proposed draws of the parameters are taken from a Gaussian proposal distribution $\pi$ with mean equal to the previous accepted draw. For a general parameter $\beta$ and iteration $p$ of the MCMC, we draw a proposed value of $\beta^{'} \sim N(\beta^{p-1}, \sigma^2)$ based on using the previous iteration $\beta^{p-1}$. The acceptance ratio is calculated as $\frac{P(\beta^{'})}{P(\beta^{p-1})}\times \frac{\pi(\beta^{'})}{\pi(\beta^{p-1})}$ where $P(\beta)$ represents the posterior distribution of $\beta$ and $\pi$ represents the proposal density. For a general Gaussian density, $g(\beta^{'}|\beta^{p-1})=$ $\frac{1}{\sqrt{2\pi \sigma^2}} \exp(-1/2\sigma^2) (\beta^{'} -\beta^{p-1})^2 $ and $g(\beta^{p-1}|\beta^{'})=$ $\frac{1}{\sqrt{2\pi \sigma^2}} \exp(-1/2\sigma^2) (\beta^{p-1} -\beta^{'})^2 $. Based on our proposal distribution, the exponential terms in the ratio of Gaussian densities will cancel, so the proposed draw $\beta^{'}$ is accepted with the simplified probability $min(1, \frac{P(\beta^{'})}{P(\beta^{p-1})})$. The variance of the proposal distribution $\sigma^2$ is tuned to obtain convergence of parameter draws and target a reasonable acceptance rate (Gelman et al. 1996).

The frailties are also drawn using a Metropolis-Hastings step with a Gaussian proposal distribution with mean equal to the previous value and a Gaussian prior with mean zero and standard deviation equal to 0.4. Each proposed frailty term for an individual has its own acceptance ratio. For $i=1,...,\frac{n}{2}$, we obtain draws of $\omega_{12i}^{0}, \omega_{13i}^{0}$, and for $i= \frac{n}{2}+1,...,n$, we obtain draws of $\omega_{12i}^{1}, \omega_{13i}^{1}$ using the posterior distribution. When we do not make assumptions of frailties being equal ($\omega_{13i}^z = \omega_{23i}^z)$, we must estimate the set $\omega_{12i}^z, \omega_{13i}^z, \omega_{23i}^z$ for each individual. $T_{23i}^z$ and corresponding $\omega_{23i}^z$ does not exist for any individual that does not experience the intermediate event. In this case, $\omega_{23}$ can be drawn directly from the prior or its conditional multivariate normal distribution in Section 2.1 using our model formulation with six frailties and a fixed covariance matrix.

\begin{figure}[H]
 \centering

\includegraphics[width = 5.9in]{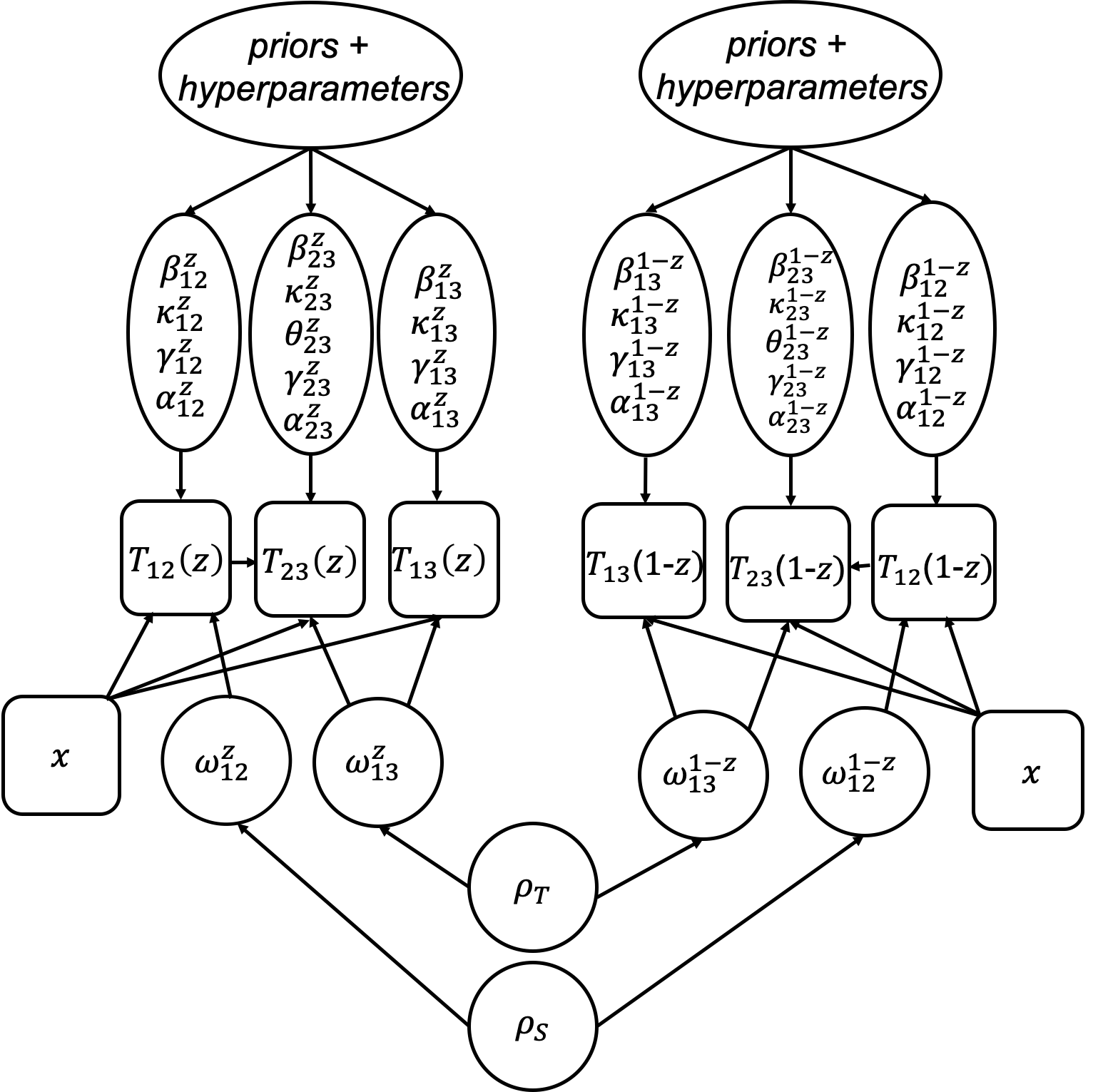}
 \caption{This diagram demonstrates the relationships between the parameters and data in the proposed model (model A assuming that $\omega_{13}^z = \omega_{23}^z$).}
\label{fig:parameters}
\end{figure}

The likelihood contributions for $L$ for each parameter can be found in an appendix. Based on the given likelihood components and prior distributions $\pi^{*}$, the posterior $P$ for a given $Z = z$ is the product over individuals $i$ who received $z$:

\noindent $$\resizebox{\textwidth}{!}{
$\prod_i \big(L_i(T_{13i}(z), T_{23i}(z), T_{12i}(z), \delta_{Si}, \delta_{Ti}; \omega_{12i}^z, \omega_{13i}^z, \omega_{23i}^z, \beta_{12}^z, \gamma_{12}^z, \alpha_{12}^z, \beta_{13}^z, \gamma_{13}^z, \alpha_{13}^z, \beta_{23}^z, \gamma_{23}^z, \alpha_{23}^z, \theta_{23}^z, \kappa_{12}^z, \kappa_{13}^z, \kappa_{23}^z) \times$
}$$
$$\resizebox{\textwidth}{!}{
$\pi^{*}(\omega_{12i}^z,\omega_{13i}^z,\omega_{23i}^z)\big) \pi^{*}(\beta_{12}^z) \pi^{*}(\gamma_{12}^z) \pi^{*}(\alpha_{12}^z) \pi^{*}(\beta_{13}^z) \pi^{*}(\gamma_{13}^z) \pi^{*}(\alpha_{13}^z) \pi^{*}(\beta_{23}^z) \pi^{*}(\gamma_{23}^z) \pi^{*}(\alpha_{23}^z) \pi^{*}(\theta_{23}^z)\pi^{*}(\kappa_{12}^z)\pi^{*}(\kappa_{13}^z)\pi^{*}(\kappa_{23}^z) $
}$$
Visually, we can see the hierarchy of parameters across different treatments and transitions and how the terms are related in Figure \ref{fig:parameters}.

Initial estimates of the frailties may be calculated using the \texttt{frailtypack} or \texttt{frailtyEM} packages in \texttt{R} (R Core Team; Rondeau and Gonzalez, 2005; Balan and Putter, 2019). Parameter estimates are each drawn from the proposal distribution individually. Under the parameterization in model A, $\theta_{23}^z$ is drawn from a proposal distribution with a mean based on the estimated coefficient from a hazard model fit using observed data regressing time to $T$ on time to $S$, among those who experience $S$. By doing this, $\theta_{23}^1$ and $\theta_{23}^0$ have unique starting values. The draws are accepted in blocks for the Metropolis-Hastings step. The blocks are divided into treatment arm transitions, and the parameters within a block are jointly accepted or rejected. For model A, we have blocks $\omega_{12i}^0; \{ \gamma_{12}^0, \alpha_{12}^0 \};\omega_{13i}^0;\{ \gamma_{13}^0, \alpha_{13}^0 \}; \{ \gamma_{23}^0, \alpha_{23}^0, \theta_{23}^0, \kappa_{23}^0 \};$ $\omega_{12i}^1 ;$ $\{ \gamma_{12}^1, \alpha_{12}^1 \} ;\omega_{13i}^1;\{ \gamma_{13}^1, \alpha_{13}^1 \};$ $\{ \gamma_{23}^1, \alpha_{23}^1, \theta_{23}^1, \kappa_{23}^1\}$ when all of the model parameters are being estimated. The proposal distributions have standard deviation $\sigma = 0.1$. 

\section{CEP Quantities}

We develop a method for validating a surrogate endpoint using the principal stratification framework (Frangakis and Rubin, 2002). The goal of this validation procedure is to develop causal quantities that rigorously determine if a time-to-event $S$ is a valid surrogate for use in a future trial in place of $T$ by conditioning on the joint distribution of the observed and counterfactual of $S$, specifically the log cumulative hazard ratio of the time to $S$ under control versus treatment. In a non-survival setting, Gilbert and Hudgens (2008) define a principal surrogate endpoint for a binary $T$ based on the comparison of the quantities $risk_{(1)}(s_1, s_0) \equiv P(T(1) = 1| S(1) = s_1, S(0) = s_0)$ and $risk_{(0)}(s_1, s_0) \equiv P(T(0) = 1| S(1) = s_1, S(0) = s_0)$. The condition that these must be equal for all $s_1 = s_0$ is known as average causal necessity. Average causal sufficiency is defined as $risk_{(1)}(s_1, s_0) \neq risk_{(0)}(s_1, s_0)$ for all $|s_1-s_0|>C$ for some non-negative constant $C$. They define the causal effect of the treatment on the true endpoints as $h(P(T(1) = 1), P(T(0) = 1))$ for some $h(,)$ contrast function that satisfies $h(x,y)=0$ if and only if $x = y$. The CEP surface is therefore equal to $h(risk_{(1)}, risk_{(0)})$ over values of $s = (s_1, s_0)$. A specific case of this is the CEP plot of $\Delta T = E(T(1) - T(0) | S(1) - S(0) = s)$ over values of $ \Delta S = S(1) - S(0) = s$ when $S$ and $T$ are continuous. Based on these criteria, an ideal CEP plot for a valid surrogate will go through the origin and have a positive slope. We generalize this by defining new contrasts, $\Delta T_i$ and $\Delta S_i$ for each subject in this time-to-event setting, forming a scatterplot of $(\Delta S_i, \Delta T_i)$, and assessing whether a straight line through the points on this scatterplot goes through the origin and has a positive slope. For $\Delta T_i$ we will use $P(T_i(1) > \tau_T) - P(T_i(0) > \tau_T)$ evaluated at time $\tau_T$. For $\Delta S_i$ we will use $\log \left( \frac{\Lambda_{12i}^0(\tau_S)}{\Lambda_{12i}^1(\tau_S)} \right)$ that depends on some time $\tau_S$. Since the intercept and the slope of the line depend on $\tau_S$ and $\tau_T$ we can write the line as $\Delta T_i (\tau_T)=\gamma(\tau_S,\tau_T)_0+
\gamma(\tau_S,\tau_T)_1 \Delta S_i (\tau_S)$. A good surrogate will have $\gamma(\tau_S,\tau_T)_0=0$ and $\gamma(\tau_S,\tau_T)_1>0$, with larger values of $\gamma(\tau_S,\tau_T)_1$ implying better surrogacy. Furthermore, for the surrogate to be relevant we would want a treatment effect on $S$, so from the CEP plot we would also assess whether the mean of $\Delta S_i$ is equal to zero.

$\tau_S$ and $\tau_T$ must be chosen at meaningful or sensible times. $\tau_T$ would usually be determined by the clinical context, and $\tau_S$ needs to be less than $\tau_T$ for the surrogate to be useful. While small times for $\tau_S$ and $\tau_T$ are desirable they should also be chosen such that a sufficient number of events have occurred in order to make sensible decisions about the surrogate. It is also possible to use the other quantities for both $\Delta S$ and $\Delta T$. Here we have chosen this $\Delta T$ as an interpretable quantity that might be used as the true endpoint in the trial, that can be calculated regardless of whether $S$ has occurred.
We have chosen $\Delta S$ to be directly related to the transition from state 1 to state 2 in the illness death model, as this is what the therapies are usually aiming to modify. Other choices for $\Delta S$ are possible in which it is based on a probability rather than a cumulative hazard or involves more than just the transition from state 1 to state 2. These will be considered in the discussion section.

While counterfactual draws of the frailties are not needed for the estimation procedure, they are needed to form the proposed CEP plot. As the correlations between the observed and counterfactual outcomes are non-identified, we fix $\rho_S, \rho_T$ from the distributions in equation 4 to draw the counterfactual frailty terms. We use correlations of 0.5 as a starting point since it is a mid-point between perfect and no correlation and then vary $\rho_S$ and $\rho_T$ for sensitivity analysis. We use the normal prior distribution and fixed $\rho_S, \rho_T$ to obtain draws of the $\omega$ estimates in the counterfactual arm from the appropriate conditional normal distributions, such as
$\omega_{12}^{z} | \omega_{12}^{1-z} \sim N(0 + \rho_S(\omega_{12}^{1-z}), 1-\rho_S^2)$ and similarly $\omega_{13}^{z} | \omega_{13}^{1-z} \sim N(0 + \rho_T(\omega_{13}^{1-z}), 1-\rho_T^2)$. We repeat the process for the other treatment arm to obtain sets of counterfactual frailties for each individual.

Each individual has a set of subject-specific hazards that will be used in a CEP plot. Let $\Delta S_i = \log \frac{{\Lambda}_{12}^0 (\tau_S| \omega_{12i}^0, x_i)}{{\Lambda}_{12}^1 (\tau_S| \omega_{12i}^1, x_i)}$ be on the x-axis of the plot where $ {\Lambda}_{12}^0 (\tau_S| \omega_{12}^0, x) = \int_0 ^{\tau_S} {\lambda}_{12}^0 (t| \omega_{12}^0, x)dt$ and $ {\Lambda}_{12}^1 (\tau_S| \omega_{12}^1, x) = \int_0 ^{\tau_S} {\lambda}_{12}^1 (t| \omega_{12}^1, x)dt$. For the y-axis, consider $\Delta T_i = P(T_i(1) > \tau_T| \omega_{12i}^1, \omega_{13i}^1, \omega_{23i}^1, x_i) -$ $ P(T_i(0) > \tau_T| \omega_{12i}^0, \omega_{13i}^0, \omega_{23i}^0, x_i)$ based on the frailties in model A. For example, using model A, $\Delta S_i = \log \frac{\Lambda_{12,0}^0(t)\exp(\kappa_{12}^0 \omega_{12i}^0)}{\Lambda_{12,0}^1(t)\exp(\kappa_{12}^1 \omega_{12i}^1)}$ if baseline covariates are not included.

Overall survival at time $\tau$ can be decomposed into components based on $ P($do not experience $S$ or $T$) + $P($experience $S$ but not $T$). More formally, this framework is similar to the likelihood for a joint illness-death model developed in Suresh et al. (2017) and for illness-death with a cure fraction proposed by Conlon et al. (2014b) and Beesley et al. (2019). In formal notation, we are interested in the quantities
$$P(T(0)>\tau_T) = P(T(0)>\tau_T,S(0)>\tau_T) + P(T(0)>\tau_T,S(0)<\tau_T)$$
and $$P(T(1)>\tau_T) = P(T(1)>\tau_T,S(1)>\tau_T) + P(T(1)>\tau_T,S(1)<\tau_T)$$
These quantities can be written in terms of parameters\\
\resizebox{\textwidth}{!}{$\exp(-\int_0^{\tau_T} \lambda_{12}(u)du - \int_0^{\tau_T}\lambda_{13}(u) du) + \int_0^{\tau_T} \exp(-\int_0^u \lambda_{12} (v)dv - \int_0^{u} \lambda_{13}(v)dv) \lambda_{12}(u)\exp(-\int_0^{\tau_T-u} \lambda_{23}(v|u)dv)du$}
$$=\exp(-\Lambda_{12}(\tau_T) - \Lambda_{13}(\tau_T)) + \int_0^{\tau_T} \exp(-\Lambda_{12}(u) - \Lambda_{13}(u)) \lambda_{12}(u)\exp(-\int_0^{\tau_T-u} \lambda_{23}(v|u)dv)du$$

Based on the draws of model parameters for a given iteration of the MCMC, we estimate observed and counterfactual hazards for each individual. After calculating $\Delta{T_i}$ and $\Delta{S_i}$ conditional on the set of $\omega_i$, we create a scatterplot of $\Delta T_i$ vs. $\Delta S_i$ and draw a loess or linear curve through the points for a single iteration of the algorithm. Our $\gamma_0$ and $\gamma_1$ summary quantities are equal to the intercept and slope of this line (whereas these quantities may need to be redefined for a loess curve). This process is repeated for the next set of draws of model parameters and frailties for all individuals. These quantities are then averaged over MCMC iterations after a burn-in period.
 
\subsection{Valid Surrogates under an Illness-Death CEP Curve}
 
As our CEP curve is a fairly complex function of the parameters and frailties, we empirically investigate what combination of illness-death models, meaning relationship between $S$ and $T$, leads to CEP plots that align with an intuitive notion of whether $S$ is a good surrogate for $T$. We primarily consider the eight scenarios that may exist based on which transitions have treatment effects (defined as whether or not the counterfactual hazards are equal) in Table \ref{tab:sim3a} and in an appendix. These scenarios and the magnitude of the effects determine whether there are marginal treatment effects on $S$ and $T$.
 
We characterize the CEP curves under these scenarios using true generating parameter values to calculate $\Delta T$ and $\Delta S$. In an appendix, we show scatterplots of $\Delta S_i$ vs. $\Delta T_i$ for simulated data, for which the values of the frailties are known. An Rshiny app is also available at \url{https://emilyroberts.shinyapps.io/id_cep_parameters/} that allows users to characterize the CEP curve for different parameter values. We also allow for the user to vary which independence or equivalence assumptions are made about the frailty terms and the corresponding impact on the CEP curve.

Based on several settings investigated in an appendix, we suggest which data scenarios should correspond to a decision that the intermediate outcome is in fact a valid surrogate. We identify that for a perfect surrogate, the paths that treatment effects should exist are through the baseline to intermediate outcome transition only (ie $\lambda_{12}^0 \neq \lambda_{12}^1$). In the null case with no treatment effects, Scenario 1, and this ideal case Scenario 2, the estimated slope is positive, and the intercept is equal to 0. This is consistent with our consideration of the more flexible Prentice Criteria, which also suggest that hazards from baseline to $S$ should be non-equal ($\lambda_{12}^0 \neq \lambda_{12}^1$) and the hazards from baseline to $T$ should be equal ($\lambda_{13}^0 = \lambda_{13}^1$) across treatment arms. Largely, small changes in the values of $\rho$ in the correlation matrix of the frailty terms does not have a major impact on the CEP slope and intercepts, though other settings in the online app demonstrate specific settings where these correlations may be more consequential.

We can examine the marginal effects on $S$ and $T$ based on the average of $\Delta S_i$ and $\Delta T_i$ and via Kaplan Meier curves in the app and quantities in an appendix. For scenario 1, treatment effects on both outcomes are zero, which may correspond to a treatment not worth future investigation. For other scenarios, the marginal effect on $T$ is somewhat small under the parameter values we are presenting. We did observe that Scenarios 3-8 (denoted as partial and non-surrogates) produced CEP curves that did not go through the origin and therefore were invalid. We anticipated differences between perfect, partial, and non-surrogates would be easily apparent, and while the intercepts did differ, the slope does not drastically change between the different scenarios. Under the particular parameters we investigated, the slope was positive for all of the scenarios when the baseline hazard to $T$ was larger after experiencing $S$ (ie the baseline hazard $\lambda_{0,23}^z > \lambda_{0,13}^z$ so that death occurs faster after progression). In other words, the relative magnitude of the baseline hazards for transition times $T_{12}(z), T_{13}(z),$ and $T_{23}(z)$ for a given treatment arm influences the slope and intercept of a CEP curve. A possible explanation for the small differences in slope values across scenarios is that the y-axis will always be constrained between -1 and 1 since it represents a difference in two probabilities. This quantity $\Delta T_i$ on the y-axis is a relatively complex function of multiple model parameters that may not change drastically based on relatively small changes in the baseline hazards. 

\subsection{Additional Scenarios}

In addition to which hazards are moderated by treatment being considered in the eight settings above, each combination can be crossed with whether $\theta_{23}$ and $\kappa_{23}$ are zero vs. nonzero in a factorial design. We briefly considered the former and do see that incorporating non-zero values of $\theta_{23}^z$ does change the slope and intercept of the CEP curve in an appendix. While the settings in Table \ref{tab:sim3a} and the extra settings through varying $\theta_{23}$ and $\kappa_{23}$ represent a broad range, there are many other possible scenarios that could be achieved with specific choices of the parameters. 
For example, even if a treatment slows the rate of progression to the surrogate endpoint, it is possible that time to death after progression may be more rapid on the treatment arm. In our setting, that would be seen in a positive treatment effect on the transition from baseline to $S$, but a negative treatment effect from $S$ to $T$ either through increasing the baseline hazard $\lambda_{23}^1$ or a positive value of $\theta_{23}^1$. Another possibility exists where the treatment slows the rate of progression, corresponding to a positive treatment effect from baseline to $S$, however toxicities or side effects from the treatment effect cause death from other causes, affecting the baseline to $T$ transition to have a negative treatment effect. More complex study designs might allow for patients to switch to the active treatment arm after experiencing the surrogate endpoint $S$ which could be potentially incorporated into our illness death framework by reducing $\lambda_{23}^0$.

\section{Simulation Study}

\subsection{Simulation Set-up}

Here we start with a simulation setting where we assume each baseline hazard follows a Weibull distribution where shape parameters for the baseline hazards and frailty coefficients are equal to 1. We conduct a simulation with 200 replicated datasets and $n = 600$. Data are generated under simple settings that follow the $\theta$ parameterization shown in model A. The true values of the parameters are shown in the simulation results in the first row of the table of results. Survival times are simulated based on a Weibull baseline hazard specification (Austin, 2012). We generate treatment effects by differing the scale parameters between arms, meaning $\gamma_{jk}^1 \neq \gamma_{jk}^0$. We simulate the frailties to have mean 0 and a standard deviation of 0.4 and assume that $\omega_{13}^z = \omega_{23}^z$ in our primary results settings.

We conduct the estimation procedure described in section 3 from our eight simulation scenarios, highlighting Scenario 1 with no marginal treatment effects on either endpoint (a null setting), Scenario 2 where there is a treatment effect only on $S$ (which we label as a perfect surrogate), and scenarios 3-8 where treatment effects exist such that we do not expect $S$ to be a surrogate. Because of non-identifiability due to the close link between the baseline hazard, frailties, and coefficients associated with the frailties, we assume during estimation that all $\kappa_{jk}^z = 1$. In an appendix, we also conduct sensitivity analyses by varying the assumptions that $\omega_{12}^z \perp \omega_{13}^z$ and $\omega_{13}^z = \omega_{23}^z$. There we assume that either $\omega_{12}^z \perp \omega_{13}^z \perp \omega_{23}^z$ or that all three frailties are correlated within a given counterfactual treatment arm. In this case we assume $\rho_{T1} = \rho_{T4} = 0.95$ and $\rho_{T3} = \rho_{T2} = \rho_{ST} = \rho_T = 0.5.$ and set $\tau_S = 1$ and $\tau_T = 5$.

\subsection{Simulation Results}

In this section, we show results of the estimated model parameters as well as validation quantities, the intercept $\gamma_0$, and slope $\gamma_1$. The estimation of the $\gamma_0$ and $\gamma_1$ quantities are calculated from fitting a linear best fit line through the CEP cloud at each iteration and reporting the posterior mean of these quantities for each simulated dataset. Parameter estimates are based on the posterior means and corresponding measures of variability; the average estimated standard error (SE) and the standard deviation (SD) of the posterior means are shown for the model parameters. We run the simulations for 3,000 iterations with 900 burn in draws. In addition to trace plots of the parameter draws, we assess the empirical mean and standard deviation of the estimated frailty terms over the iterations. 

\begin{figure}[H]
\centering
\includegraphics[width = 3.2in]{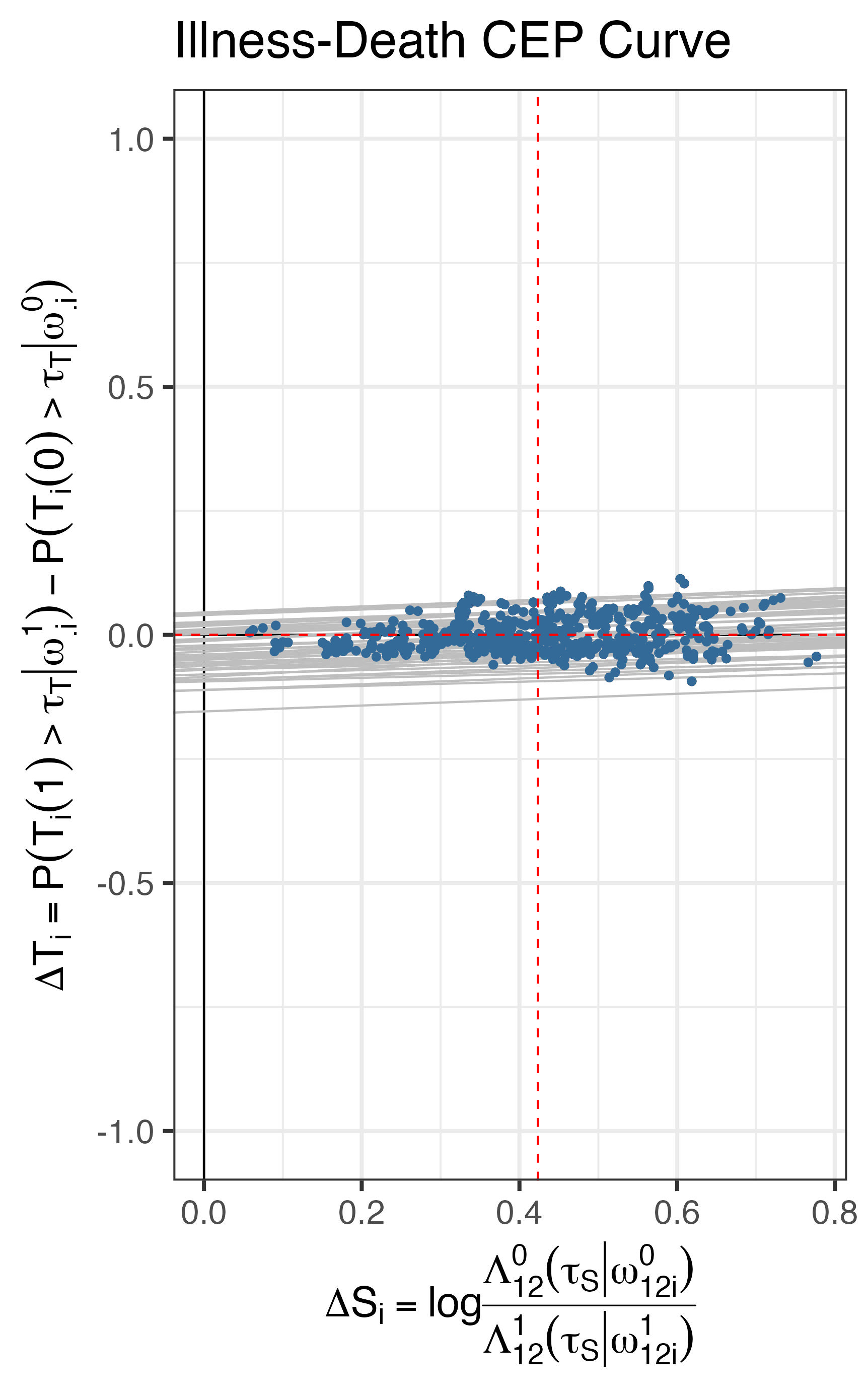}
\caption{Example of an estimated CEP curve, conditional on frailties, for a single simulated dataset under Scenario 2.} \label{fig:simCEP}
\end{figure}

In Figure \ref{fig:simCEP} we show the CEP curve conditional on estimated frailties for one dataset under Scenario 2. Each point is the posterior mean of ($\Delta S_i$, $\Delta T_i$) across MCMC iterations. The posterior values of the slope and intercept are shown, which convey the amount of variability based on the posterior coordinates of ($\Delta S_i$, $\Delta T_i$) for each individual $i$. We see that the estimated slope and intercept correctly meet our criteria of a valid surrogate under our proposed set of model assumptions. Though there is substantial variability in the estimates of $\gamma_0$ and $\gamma_1$, the respective posterior mean and credible intervals are -0.018 (-0.078, 0.042) and 0.049 (0.020, 0.078) for this dataset. Furthermore, there is a marginal effect of the treatment on both $S$ and $T$ for this dataset, as denoted by the non-zero position of the dashed lines.

In the main set of simulations in Table \ref{tab:sim3a}, the identified parameters are estimated fairly well and seem to converge based on the assumptions we have made. We observe that the distribution of the estimated frailty terms can deviate from the generating distribution with mean zero and fixed variance. {While our method involves prior and proposal distributions for the frailties,} we are not directly enforcing any assumptions about the mean or variability of the frailty parameters during the estimation algorithm. The shape of the likelihood for frailty terms, particularly $\omega_{12}^z$ terms for individuals with $\delta_{Si} = 0$, seems to be fairly flat, so the draws move around considerably during the algorithm. In these considered simulations, the credible intervals around $\gamma_1, \gamma_0$ are somewhat wide for all scenarios. Since an ideal surrogate will have values $\gamma_0 = 0$ and $\gamma_1 > 0,$ too much uncertainty can make it difficult to determine the value of the surrogate.

In our sensitivity analyses about the assumptions on the frailty terms, shown in an appendix, we see some sensitivity to the assumptions being made, such as increased variability in the subject-specific points. How these factors influence the CEP curves should be investigated under trial specific contexts.
 
\section{Prostate Cancer Example}

Our motivating clinical study is a phase III, randomized trial for men with prostate cancer, NRG/RTOG 9601 (Shipley et al., 2017). The trial features 760 men with recurrently or persistently elevated prostate-specific antigen (PSA) levels whose prostate was initially removed by prostatectomy. The two treatments being compared are post-prostatectomy radiation therapy with or without antiandrogen therapy. There are 384 and 376 men in each treatment arm. The two survival endpoints of interest are time to distant metastasis, defined as radiographic evidence of metastatic cancer, and overall survival (OS). Notably, composite endpoints such as metastasis-free survival (MFS) are often evaluated. It has been previously established by The Intermediate Clinical Endpoints in Cancer of the Prostate (ICECaP) that MFS is a valid surrogate for OS in the setting of the initial treatment for localized prostate cancer (Xie et al., 2017). Others have evaluated if MFS is a valid surrogate when assessing the impact of antiandrogen therapy in recurrent prostate cancer following post-prostatectomy salvage radiation therapy (Jackson et al., 2020). However, within our illness-death framework we consider time to distant metastasis and time to death separately. Covariates in the dataset are also available, including PSA values at the time of randomization, Gleason score, and age in grouped categories.

We show in Figure \ref{fig:cepdata} the Kaplan Meier curves for the intermediate and true outcomes without considering the semi-competing risk as well as the curve for the transition from $S$ to $T$ for those who experienced distant metastasis. $S$ may be censored because it was not observed during the study period or because the terminal event $T$ occurred first. In an appendix, in Figure A6, we also present the cumulative incidence curve for $S$ considering $T$ as a semi-competing risk based on the non-parametric Aalen-Johansen estimate of the cumulative incidence function from the \texttt{mstate} \texttt{R} package (Putter, 2011). The plots show that the addition of antiandrogen therapy decreases the hazard of distant metastases and increase the survival probability, but after metastases the survival probability is reduced and does not appear to be greatly influenced by whether the antiandrogen therapy was part of the treatment.

\subsection{Conventional Models}

We consider the $z=1$ group to be the treatment group for salvage radiation therapy with antiandrogen therapy, and the $z=0$ represents the group treated without antiandrogen therapy. There is a significant treatment effect of the additional antiandrogen therapy on time to distant metastasis using a parametric hazard model with a Weibull baseline hazard ($HR = 0.622, p = 0.004)$. There is a marginally significant treatment effect on overall survival when considering the cause-specific hazard ($HR = 0.722, p = 0.049)$. As a way to consider the Prentice criteria, we also fit a model for time to overall survival adjusting for the occurrence of distant metastases as a time-dependent covariate. We found that the effect was attenuated toward null ($HR = 0.890, p = 0.592)$ and no longer statistically significant. Based on the Kaplan Meier curves and typical survival times, we chose $\tau_S = 5$ and $\tau_T$ = 8. We calculate the number of individuals who go through each transition and experience the events in our illness-death models. In total, 156 patients experienced distant metastases, and 239 total deaths were observed between the two arms. These numbers are shown in Figure \ref{iddataset}.

\subsection{Surrogacy Evaluation} 

Here we perform the analysis marginally, without including baseline covariates. We show an estimated CEP curve based on several assumptions: the baseline hazard follows an exponential distribution, and we use model A using $T_{12}$ as a time-varying covariate where we assume $\kappa_{12}^z = \kappa_{13}^z = \kappa_{23}^z = 1$. Table \ref{tab:dataresults} shows the posterior mean and corresponding 95\% credible interval for each parameter being estimated. We plot the posterior mean of $\Delta S_i$ and $\Delta T_i$ for each individual across iterations in a CEP plot. We also show the estimated slope and intercept lines on the CEP curve for each iteration of the MCMC chain to assess the variability of the estimates of these validation quantities.

\begin{figure}[H]
\centering
\includegraphics[width = 4in]{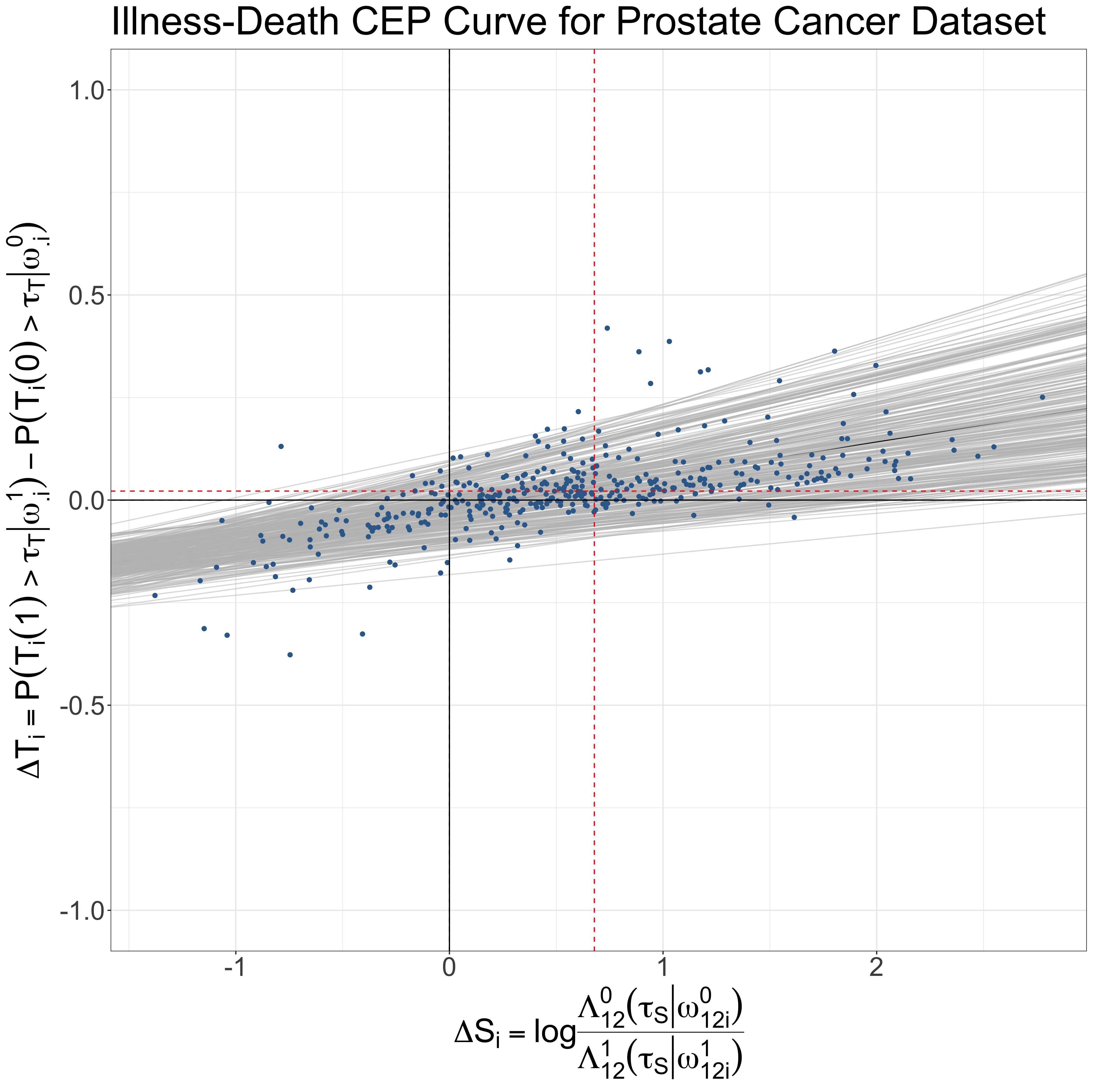}

\caption{Causal effect predictiveness plot for the motivating prostate cancer trial dataset. Each point represents the posterior mean of $\Delta S_i$ and $\Delta T_i$ for an individual. The collection of linear best fit lines in gray represent the posterior slope $\gamma_1$ and intercept $\gamma_0$ evaluated at each iteration of the MCMC. The posterior marginal effects on $S$ and $T$ are shown in the red dotted lines. No covariates are considered in this model.} \label{dataCEP}
\end{figure}

Based on this example dataset and CEP curve in Figure \ref{dataCEP}, the vertical and horizontal lines for the marginal treatment effects are separated from zero, and the posterior mean for the intercept term $\gamma_0$ is -0.036 with 95\% credible interval (-0.152, 0.080). For the slope $\gamma_1$, the posterior mean is 0.076 with 95\% credible interval is (0.017, 0.135). Based on these estimates, we would conclude that the slope $\gamma_1$ is positive, and the estimated intercept $\gamma_0$ is near zero since the credible interval for $\gamma_0$ does include 0. These results would indicate that the surrogate seems valid, though the credible interval for $\gamma_0$ is somewhat wide. We also conducted a sensitivity analysis where instead of assuming $\omega_{13}^z = \omega_{23}^z$ and that $\omega_{12}^z \perp \omega_{13}^z$, we assumed that all six counterfactual frailties were correlated within an individual. These results gave reasonably similar conclusions, with an estimated $\gamma_0$ of -0.046 (-0.157, 0.073) and estimated $\gamma_1$ of 0.108 (0.045, 0.195).

\section{Discussion and Conclusion}

In this work, we have considered how to validate surrogate endpoints when trial outcomes are time-to-event using principal stratification and illness-death models. We believe the illness-death framework is foundational to modeling these data, though a single, optimal estimand corresponding to the model is less obvious. We have provided examples and an online app to explore CEP curves under different data settings. While the values of the CEP curve can be written in a closed, analytic form when the outcomes are Gaussian in previous work (Conlon et al., 2014a; Roberts et al., 2021), it is necessary to define and empirically assess what an ideal CEP curve looks like for time-to-event data. A novel distinction in this work is that in the Gaussian case, the CEP conditions on $S_i(1)-S_i(0)=s$, where the conditioning is on a contrast between potentially observable values, $S_i(1)$ and $S_i(0)$. In this paper, we are looking at the contrast between $\Lambda_{12i}^z$ and $\Lambda_{12i}^{1-z}$, which is a contrast between distributions. 

While not the case in our considered scenarios, some extrapolation may be required to determine if the CEP curve goes through the origin of the plot depending on the size of the treatment effect on $S$. The subject-specific points may not appear in all four quadrants of the plot. There is an interesting connection regarding individual specific $\Delta S_i$ and $\Delta T_i$ within the quadrants of the graph that has been considered across trials in the meta-analytic setting (Elliott et al., 2015). In particular, certain subject-specific coordinates may suggest that the treatment has a beneficial effect on the surrogate endpoint but a detrimental effect on the true outcome for certain individuals. This may be informative when considering the possibility of the surrogate paradox (VanderWeele, 2013).

There are several areas for sensitivity analyses and exploration of identifiability for surrogacy validation (Ghosh, 2012). While the variance of the frailty should be identifiable by including sufficient covariates (Gao, 2012; Putter et al., 2015), it may still be difficult to accurately estimate frailty terms in a complex model. In our proposed models, we include a prior distribution for the variance of the frailty terms but do not assume the variance is known. Since allowing for too much flexibility in the models may result in non-identifiability of parameters, this can lead to computational problems when trying to estimate the coefficients associated with the frailties. We believe our assumptions that $\kappa_{jk}=1$ or that $\omega_{13}^z = \omega_{23}^z$ about the frailty terms are justifiable for this data example. They also help with computation during estimation, but they are still potentially strong assumptions. Relaxing the assumption that the frailties going into the $T$ state are equal (ie $\omega_{13}^z = \omega_{23}^z$) may impact identifiability since there will be less information available to estimate these terms. To the extent that frailties can be estimated for one event time per person, the data might inform these assumptions (e.g., the assumption is testable to the extent that frailties can be estimated well). We might try to assess the identifiability of frailty terms in the proposed causal model by comparing the prior and posterior distributions for the frailty terms (Gao, 2012). Other convergence metrics can be used to assess the convergence of the parameters, and more complex algorithms or different distributional assumptions about the frailties may alleviate computational problems (Clayton, 1991; Wen et al., 2016 for example). For assessment of robustness, our models can be evaluated under model misspecification. To increase the flexibility of the method, we could also consider fitting a non-linear loess curve through the points on the CEP plot as opposed to a linear fit. We can compare our proposed methods to copula models (Taylor et al., 2015). These particular Gaussian copula models have potential of extending the closed-form correlation structure we have focused on in previous work while incorporating conditional independence assumptions on the appropriate correlation scale.

We could fit conditional surrogacy validation models and include baseline PSA, age, and Gleason score as baseline covariates. It is likely that controlling for covariates will change the estimated frailties, as frailty terms capture unexplained heterogeneity in treatment effects which would then be partially explained by the covariates. Based on these analyses, we could also determine if the surrogate is valid for certain subgroups of people (Roberts et al. 2021). Different covariates may be more important in different transition models. For example, we may expect age to be more important for the direct transition from baseline to death, while baseline PSA and Gleason score will likely be more important for time to distant metastases. Model selection could lower the number of parameters to estimate (Reeder et al., 2022). 

In the future, we can consider changing our model parameterization from our proposal to use a time-varying covariate in the transition model from $S$ to $T$ to the alternative Model B or a different structure. We may extend beyond the proposed illness-death model to a different or more complex multi-state model depending on the endpoints being evaluated. In different disease areas, consideration about individuals being cured may be appropriate (Conlon et al, 2014b). We have assumed here that time to $S$ is known, but it may be subject to interval censoring (Zeng et al. 2018). In some cases we may even have exact information about time to $T$ based on death registries without knowing if $S$ occurred (Beesley et al., 2019). Different models, definitions of the endpoint, and corresponding $\Delta S_i$ may change our determination whether the surrogate is valid, and the assumptions made about the models and frailties may be more appropriate for certain contexts. 

While we believe the illness-death model is natural for modeling these data, different estimands could be considered for validation. In the CEP plot we have used the ratio of the cumulative hazards on the horizontal axis as a measure of the treatment effect on $S$. This was chosen because it is explicitly related to the transition from the baseline state to the state of experiencing the surrogate, and in most settings, including our prostate cancer one, the primary way in which the treatment is expected to work is by preventing or delaying the occurrence of the events in the intermediate state. There are other possible choices for what to use for $\Delta S$ on the horizontal axis. One would be based on the difference in the cumulative incidence of $S$ by time $\tau_S$ between the two arms, another could be based on the composite endpoint of either $S$ or $T$ occurring by time $\tau_S$. Both of these can be calculated from the illness death model parameter estimates, but both are also impacted by the transition rate from the baseline state to the terminal state.

It would be interesting to evaluate this illness death model when $\Delta S_i$ is based on a composite endpoint with $T$. For example, in the prostate cancer setting, distant metastases-free survival has been considered as a surrogate endpoint for overall survival. Other potential surrogates have been considered such as biochemical recurrence or time to local recurrence, and an alternative true clinical outcome could be prostate cancer-specific survival. In our setting, it is likely that individuals may only die from prostate cancer if they experience distant metastases, so there may be fewer individuals transitioning directly from baseline to cancer-specific death compared to baseline to death from other causes. 

In this paper we have considered the situation of a single trial, in contrast to the meta analysis setting in which data from multiple trials are analyzed. We developed an approach to assessing whether $S$ is a valid surrogate for $T$ from a causal perspective. The hope would be that if $S$ is a good surrogate for $T$ in one trial, then it would also be a good surrogate for $T$ in other trials with similar treatments. The fact that the surrogacy measure is based on causal concepts, not just measures of association, may make it more likely to transport from one trial to the next. In previous work, in a different data setting, we find similar CEP plots across four different treatments comparisons (Taylor et al., 2015). Furthermore, in this paper, a mechanistic approach to disease progression, implicit in the illness-death model, has been taken. This illness-death structure does transport from one trial to the next, so may believe that our approach will assess surrogates in a way that is more generalizable across treatments than methods that rely on composite endpoints, that do not require the illness death structure. This comparison, concept of transportability, and potential need for replication across several trials remain as future work (Pearl and Bareinboim, 2011).

There are other directions for extending this work, particularly when considering the overlap of causal inference and survival analysis and delicate interpretation of hazard ratios with multiple time-to-event endpoints. Others (Gran et al. 2015; Valeri et al. 2021) explore other causal tools for multi-state models such as inverse probability weighting, G-computation, and manipulating hypothetical transition intensities. Other directions for future work are to formally compare the proposed models with the similar structures of the Prentice criteria, models using mediation strategies, or other causal methods. 

\newpage 

\bibliographystyle{apalike}
\bibliography{reference.bib}
\nocite{*}

\newpage

\section*{Conflict of Interest Statement:} None to report.

\section*{Acknowledgements}

We would like to acknowledge NRG for the RTOG data and thoughtful feedback by Drs. Matthew Schipper, Walter Dempsey, and Ben Hansen on this work.

\section{Figures and Tables}

\begin{singlespace}

\begin{table}[H]
     \centering
     \begin{tabular}{c|cccc}
          &  $\lambda_{12}^{0} = \lambda_{12}^{1}$ & $\lambda_{13}^{0} = \lambda_{13}^{1}$ & $\lambda_{23}^{0} = \lambda_{23}^{1}$ & Surrogacy \\ \hline
                  Scenario 1  & T & T & T & Null Case \\

                  Scenario 2  & F & T & T & Perfect\\
        Scenario 3  & F & T & F & Partial\\
        Scenario 4  & F & F & T & Partial\\
        Scenario 5  & F & F & F & Partial \\
        Scenario 6  & T & F & F & Not a surrogate\\

        Scenario 7  & T & T & F & Not a surrogate\\
        Scenario 8  & T & F & T & Not a surrogate\\
     \end{tabular}
     \caption{Eight possible scenarios of which pathways in the illness death models exhibit treatment effects based on the causal hazards. $T$ denotes true and $F$ denotes false. The right hand column represents an intuitive notion of whether $S$ is a good surrogate for $T$.}
     \label{tab:factorialapp}
 \end{table}
  \end{singlespace}

\begin{singlespace}

\newgeometry{left=.7in,right=.1in,top=.9in,bottom=.9in,nohead}
\afterpage{\begin{landscape}

\begin{table}[H]
\small

 \begin{tabular}{ccccccccccccccccc}
& $\gamma_0$ & $\gamma_1$ & $\alpha_{12}^{0}$ & $\alpha_{13}^{0}$ & $\alpha_{23}^{0}$ & $\alpha_{12}^{1}$ & $\alpha_{13}^{1}$ & $\alpha_{23}^{1}$ & $\gamma_{12}^{0}$ & $\gamma_{13}^{0}$ & $\gamma_{23}^{0}$ & $\gamma_{12}^{1}$ & $\gamma_{13}^{1}$ & $\gamma_{23}^{1}$ & $\theta_{23}^0$ & $\theta_{23}^1$ \\
\hline 1: True Value & -0.001 & 0.058 & 1 & 1 & 1 & 1 & 1 & 1 & 1 &0.5&1& 1 & 0.5 & 1 & 0 & 0 \\ \hline
Estimates & -0.006 & 0.051 & 1.030 & 1.013 & 1.064 & 1.030 & 1.023 & 1.074 & 0.991 & 0.487 & 0.881 & 1.011 & 0.508 & 0.889 & 0.016 & -0.001 \\ 
  SE & 0.047 & 0.014 & 0.039 & 0.054 & 0.058 & 0.038 & 0.069 & 0.064 & 0.069 & 0.049 & 0.082 & 0.071 & 0.075 & 0.109 & 0.084 & 0.090 \\ 
  SD & 0.042 & 0.012 & 0.029 & 0.040 & 0.047 & 0.031 & 0.037 & 0.048 & 0.086 & 0.051 & 0.085 & 0.100 & 0.053 & 0.091 & 0.085 & 0.069 \\ 
\hline
2: True Value & -0.001 & 0.058 & 1 & 1 & 1 & 1 & 1 & 1 & 1 &0.5&1& 0.61 & 0.5 & 1 & 0 & 0 \\ \hline
Estimates &  0.003 & 0.051 & 1.025 & 1.015 & 1.069 & 1.032 & 1.031 & 1.071 & 1.002 & 0.497 & 0.873 & 0.611 & 0.490 & 0.891 & 0.019 & -0.015 \\ 
  SE & 0.048 & 0.014 & 0.038 & 0.053 & 0.059 & 0.050 & 0.065 & 0.073 & 0.070 & 0.050 & 0.081 & 0.047 & 0.055 & 0.135 & 0.084 & 0.084 \\ 
  SD & 0.041 & 0.011 & 0.026 & 0.039 & 0.055 & 0.042 & 0.045 & 0.056 & 0.086 & 0.051 & 0.088 & 0.062 & 0.049 & 0.107 & 0.087 & 0.067 \\ 
  \hline
3: True Value & 0.080 & 0.038 & 1 & 1 & 1 & 1 & 1 & 1 & 1 &0.5&1& 0.61 & 0.5 & 0.61 & 0 & 0 \\ \hline
Estimates &   0.119 & 0.020 & 1.030 & 1.016 & 1.063 & 1.036 & 1.033 & 1.086 & 0.989 & 0.495 & 0.869 & 0.614 & 0.487 & 0.492 & 0.020 & 0.036 \\ 
  SE & 0.050 & 0.014 & 0.039 & 0.054 & 0.058 & 0.050 & 0.065 & 0.076 & 0.069 & 0.050 & 0.080 & 0.047 & 0.055 & 0.078 & 0.083 & 0.085 \\ 
  SD & 0.046 & 0.014 & 0.029 & 0.037 & 0.058 & 0.045 & 0.049 & 0.069 & 0.082 & 0.051 & 0.085 & 0.063 & 0.043 & 0.072 & 0.082 & 0.076 \\ 
  \hline
4: True Value & 0.062 & 0.083 & 1 & 1 & 1 & 1 & 1 & 1 & 1 &0.5&1& 0.61 & 0.31 & 1 & 0 & 0 \\ \hline
Estimates & 0.068 & 0.076 & 1.027 & 1.013 & 1.075 & 1.047 & 1.039 & 1.074 & 0.993 & 0.493 & 0.865 & 0.604 & 0.303 & 0.914 & 0.029 & -0.025 \\ 
  SE & 0.052 & 0.014 & 0.039 & 0.054 & 0.059 & 0.051 & 0.075 & 0.072 & 0.069 & 0.050 & 0.081 & 0.043 & 0.034 & 0.148 & 0.085 & 0.076 \\ 
  SD & 0.042 & 0.011 & 0.031 & 0.035 & 0.057 & 0.053 & 0.061 & 0.051 & 0.086 & 0.050 & 0.086 & 0.052 & 0.033 & 0.101 & 0.092 & 0.045 \\ 
\hline
5: True Value & 0.153 & 0.061 & 1 & 1 & 1 & 1 & 1 & 1 & 1 &0.5&1& 0.61 & 0.31 & 0.61 & 0 & 0 \\ \hline
Estimates &   0.195 & 0.045 & 1.032 & 1.019 & 1.073 & 1.044 & 1.025 & 1.096 & 0.994 & 0.490 & 0.866 & 0.609 & 0.305 & 0.509 & 0.021 & 0.022 \\ 
  SE & 0.054 & 0.015 & 0.039 & 0.054 & 0.059 & 0.051 & 0.075 & 0.076 & 0.069 & 0.049 & 0.080 & 0.043 & 0.035 & 0.087 & 0.082 & 0.075 \\ 
  SD & 0.050 & 0.012 & 0.029 & 0.038 & 0.057 & 0.053 & 0.063 & 0.076 & 0.091 & 0.051 & 0.096 & 0.053 & 0.034 & 0.068 & 0.081 & 0.059 \\ 
  \hline
6: True Value & 0.151 & 0.061 & 1 & 1 & 1 & 1 & 1 & 1 & 1 &0.5&1& 1 & 0.31 & 0.61 & 0 & 0 \\ \hline
Estimates & 0.192 & 0.044 & 1.026 & 1.014 & 1.071 & 1.038 & 1.023 & 1.090 & 0.990 & 0.492 & 0.860 & 0.985 & 0.314 & 0.508 & 0.026 & 0.055 \\ 
  SE & 0.052 & 0.014 & 0.039 & 0.053 & 0.058 & 0.040 & 0.080 & 0.065 & 0.069 & 0.050 & 0.080 & 0.065 & 0.047 & 0.067 & 0.084 & 0.084 \\ 
  SD & 0.055 & 0.013 & 0.026 & 0.039 & 0.056 & 0.034 & 0.055 & 0.062 & 0.084 & 0.053 & 0.095 & 0.071 & 0.041 & 0.064 & 0.089 & 0.068 \\ 
  \hline
7: True Value & 0.090 & 0.038 & 1 & 1 & 1 & 1 & 1 & 1 & 1 &0.5& 1 & 1 & 0.5 & 0.61 & 0 & 0 \\ \hline
Estimates &  0.127 & 0.020 & 1.029 & 1.013 & 1.074 & 1.032 & 1.022 & 1.080 & 0.993 & 0.493 & 0.871 & 1.002 & 0.503 & 0.500 & 0.025 & 0.060 \\ 
  SE & 0.050 & 0.014 & 0.039 & 0.053 & 0.059 & 0.039 & 0.069 & 0.066 & 0.069 & 0.050 & 0.082 & 0.071 & 0.075 & 0.064 & 0.085 & 0.092 \\ 
  SD & 0.051 & 0.012 & 0.028 & 0.037 & 0.054 & 0.030 & 0.039 & 0.066 & 0.089 & 0.056 & 0.092 & 0.081 & 0.054 & 0.062 & 0.092 & 0.087 \\ 
  \hline
8: True Value & 0.051 & 0.081 & 1 & 1 & 1 & 1 & 1 & 1 & 1 &0.5 &1& 1 & 0.31 & 1 & 0 & 0 \\ \hline
Estimates &  0.043 & 0.075 & 1.031 & 1.017 & 1.069 & 1.037 & 1.027 & 1.073 & 0.992 & 0.490 & 0.874 & 0.986 & 0.314 & 0.916 & 0.023 & -0.009 \\ 
  SE & 0.050 & 0.014 & 0.039 & 0.054 & 0.059 & 0.040 & 0.080 & 0.063 & 0.069 & 0.049 & 0.081 & 0.065 & 0.047 & 0.117 & 0.083 & 0.084 \\ 
  SD & 0.044 & 0.012 & 0.027 & 0.040 & 0.053 & 0.031 & 0.051 & 0.043 & 0.074 & 0.058 & 0.101 & 0.077 & 0.039 & 0.081 & 0.090 & 0.058 \\
\hline
 \end{tabular}

 \caption{Eight possible scenarios of which pathways in the illness-death models exhibit treatment effects. Simulation results from illness-death models. This table shows the posterior mean, average estimated standard error (SE), and the standard deviation (SD) of the posterior means across simulation replications. }
 \label{tab:sim3a}

\end{table}
\end{landscape}}

\restoregeometry 

\end{singlespace}
\newpage

\newgeometry{left=1in,right=1in,top=.2in,bottom=.6in,nohead}

\begin{figure}[H]
\begin{tikzpicture}
\hspace{-.6cm}

 \tikzstyle{leaf}=[shape=circle,draw=black,fill=gray!20,minimum size=0.01cm]
	\node[leaf] (p3) at ( 2, -2) {T(0)};
 	\node[leaf] (p1) at (-3, 0) {Baseline};
 \node[leaf] (p2) at ( 2, 2) {S(0)};
 \begin{scope}[every path/.style={->}]
 		\draw (p1) to node[midway, above, sloped] {$T_{12}(0)$: 93} (p2);
 \draw[dotted] (p1) to node[midway,below, sloped] {$T_{13}(0)$: 77} (p3) ;
 \draw (p2) to node[right,midway] {$T_{23}(0)$: 54}(p3) ;
 \end{scope} 

 \tikzstyle{leaf}=[shape=circle,draw=black,fill=gray!20,minimum size=0.01cm]
	\node[leaf] (p3) at ( 12, -2) {T(1)};
 	\node[leaf] (p1) at (7, 0) {Baseline};
 \node[leaf] (p2) at ( 12, 2) {S(1)};
 \begin{scope}[every path/.style={->}]
 		\draw (p1) to node[midway, above, sloped] {$T_{12}(1)$: 63} (p2);
 \draw[dotted] (p1) to node[midway,below, sloped] {$T_{13}(1)$: 74} (p3) ;
 \draw (p2) to node[right,midway] {$T_{23}(1)$: 34}(p3) ;
 \end{scope} 
\end{tikzpicture}
 \caption{Counterfactual Illness-Death Models for baseline, illness ($S$), and death ($T$) with the number of individuals experiencing the events in each transition for the prostate cancer trial.} \label{iddataset}
\end{figure}
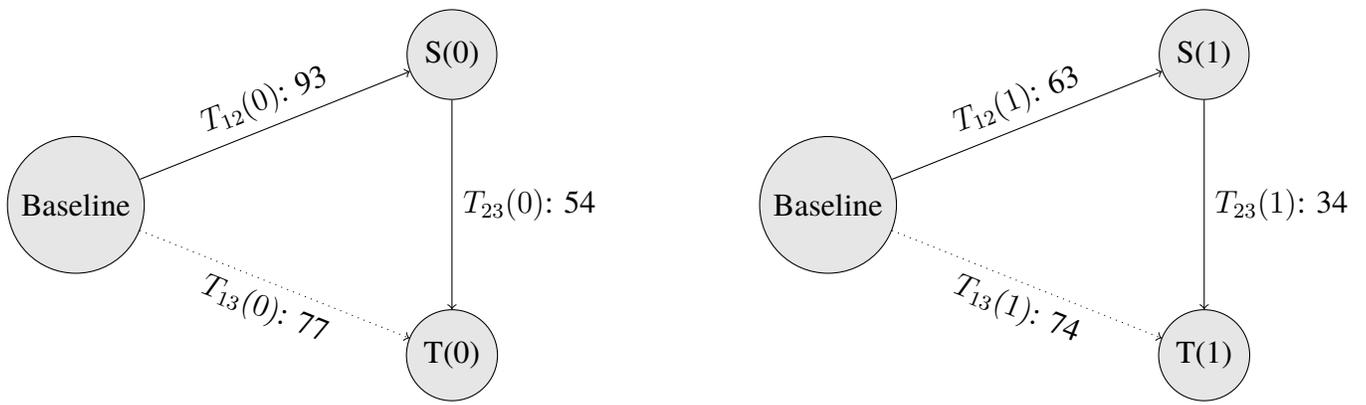
\begin{figure}[H] 
 \centering
\includegraphics[width = 3.23in]{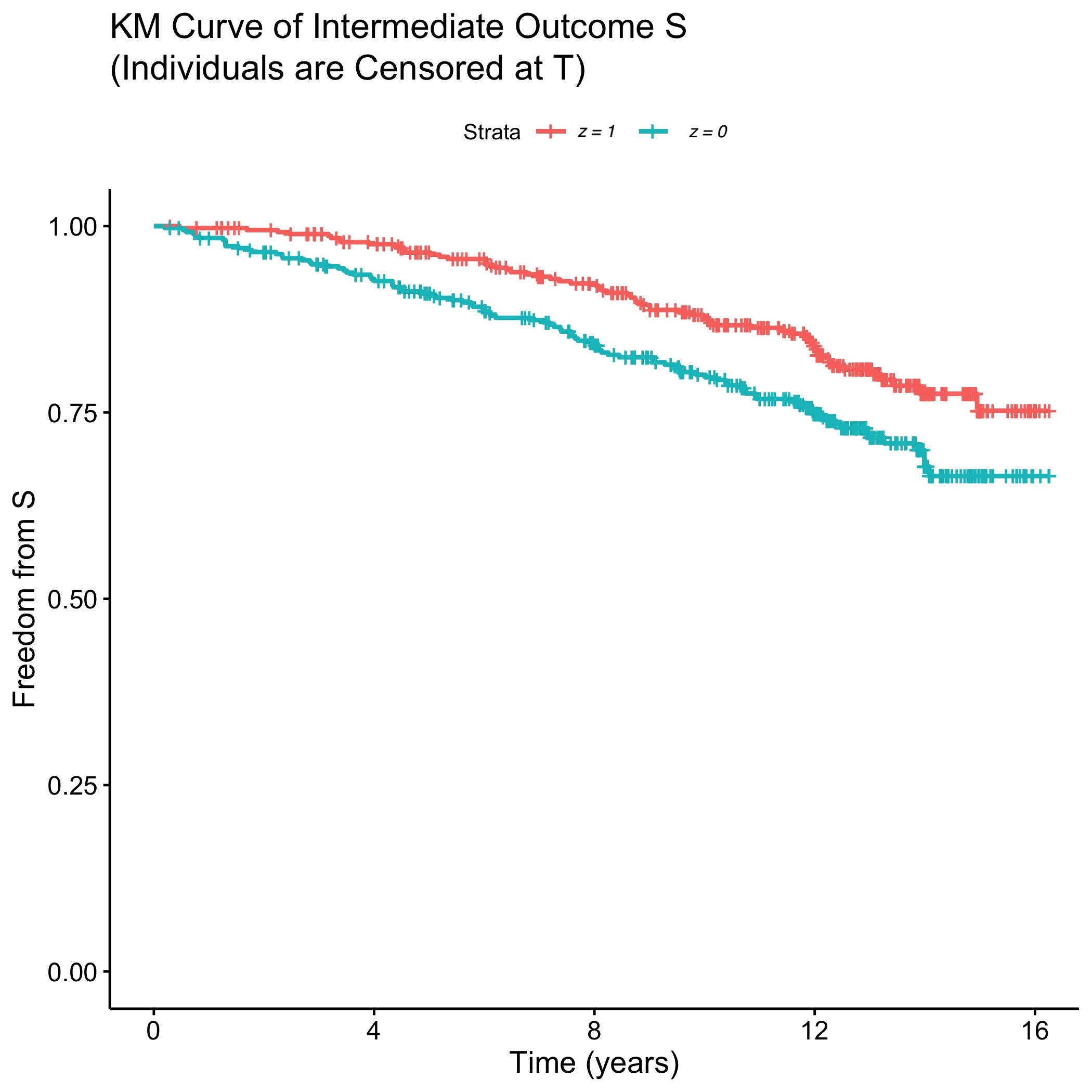} \includegraphics[width = 3.23in]{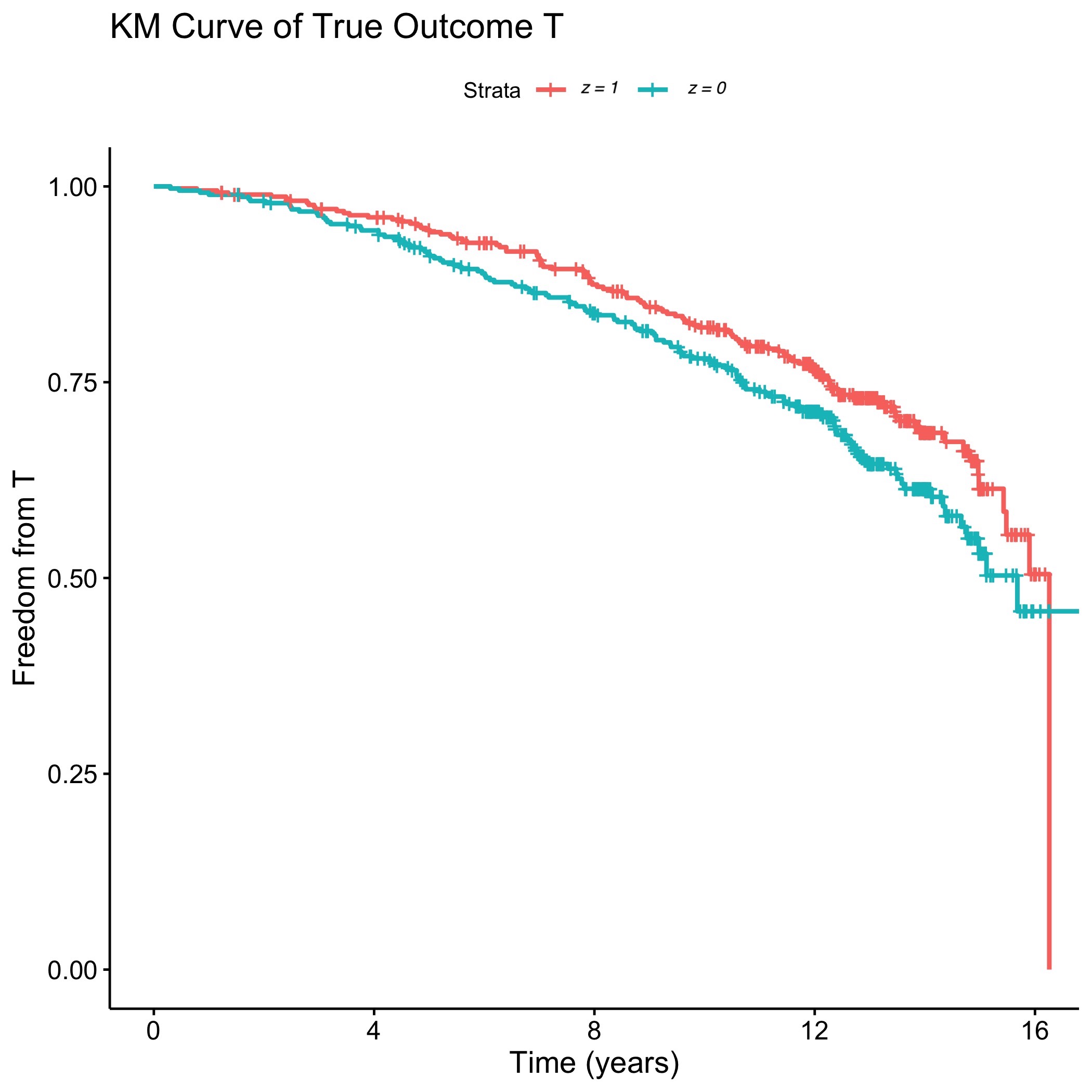} 
\includegraphics[width = 3.3in]{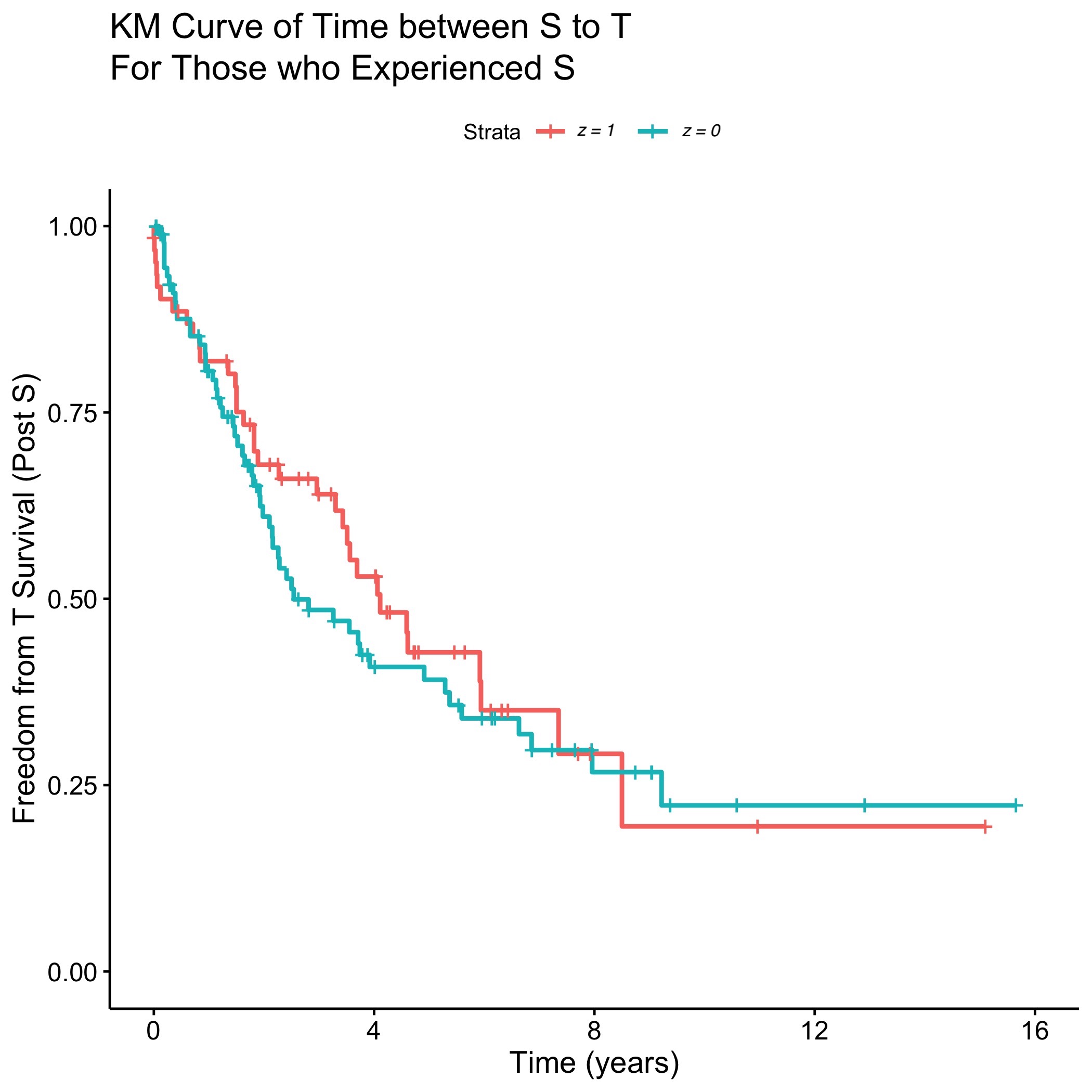} \caption{Kaplan Meier curves for the intermediate and true outcome demonstrating significant treatment effects for the prostate cancer trial. We also show the Kaplan Meier curve for the transition from $S$ to $T$ among those who experienced $S$.}
\label{fig:cepdata}
\end{figure}

\restoregeometry

\begin{singlespace}

\begin{table}[H]
 \centering
 \resizebox{\textwidth}{!}
{ \begin{tabular}{ccccccccccc}
 Parameter & $\gamma_0$ &
 $\gamma_1$ & 
 $\gamma_{12}^0$ & 
 $\gamma_{13}^0$ & \
 $\gamma_{23}^0$ & 
 $\gamma_{12}^1$ & 
 $\gamma_{13}^1$ & 
 $\gamma_{23}^1$ & 
 $\theta_{23}^0$ & 
 $\theta_{23}^1$ \\ \hline
 Posterior Mean & -0.036 & 0.076 & 0.018 & 0.018 & 0.172 & 0.013 & 0.015 & 0.266 & 0.097 & 0.035 \\
 
 SE & 0.059 & 0.030 & 0.002 & 0.002 & 0.180 & 0.002 & 0.002 & 0.371 & 0.248 & 0.243 \\
 \\

 \end{tabular}}
 \caption{Parameter estimates for the prostate cancer data example. The posterior mean and estimated standard error are shown for each parameter. All $\alpha_{jk}$ and $\kappa_{jk}$ are set to 1.}
 \label{tab:dataresults}
\end{table}

\end{singlespace}

\end{document}